\documentclass[graphicx,amsmath,preprint,unsortedaddress,showpacs,eqsecnum]{revtex4}

\usepackage{graphicx}
\usepackage{dcolumn}
\usepackage{bm}

\usepackage{float}

\usepackage{color}

\begin{document}


\title{Single versus double bond breakage in a Morse 
chain under tension: higher index saddles and bond healing} 



\author{F. A. L. Maugui\`{e}re}
\email{Frederic.Mauguiere@bristol.ac.uk}
\affiliation{School of Mathematics, University of Bristol, Bristol BS8 1TW, United Kingdom}

\author{P. Collins}
\affiliation{School of Mathematics, University of Bristol, Bristol BS8 1TW, United Kingdom}

\author{G. S. Ezra}
\affiliation{Department of Chemistry and Chemical Biology, Baker Laboratory, 
Cornell University, Ithaca, NY 14853, United States}

\author{S. Wiggins}
\affiliation{School of Mathematics, University of Bristol, Bristol BS8 1TW, United Kingdom}


\date{\today}

\begin{abstract}
We  investigate the  fragmentation dynamics of an atomic chain under tensile stress. 
We have classified the location, stability type (indices) and energy of all 
equilibria for the  general $n$-particle chain, and have highlighted 
the importance of  saddle points with index $> 1$. 

We show that for an $n=2$-particle chain under tensile stress the index 2 
saddle plays a central role in organizing the dynamics. We apply normal 
form theory to  analyze phase space structure and dynamics in a neighborhood 
of the index 2 saddle. We define a phase dividing surface (DS) that enables us 
to  classify trajectories passing through a neighborhood of the  saddle 
point using the values of the integrals associated with the normal form.  
We  also generalize our definition of the dividing surface and define 
an \emph{extended dividing surface} (EDS), which is used to sample and classify  
all trajectories that pass through a phase space neighborhood of the index 2 saddle 
at total energies less than that of the saddle.

Classical trajectory simulations are used to study single versus double bond 
breakage for the $n=2$ chain under tension.   Initial conditions for 
trajectories are obtained by sampling the EDS at constant energy. 
We sample trajectories at fixed energies both above and below the energy of the saddle. 
The fate of trajectories (single versus double bond breakage) 
is explored as a function of the location of the initial condition on the EDS, 
and a connection made to the work of Chesnavich on collision-induced dissociation.
A significant finding is that we can readily identify trajectories that exhibit
bond \emph{healing}. Such trajectories pass outside the nominal (index 1) transition state 
for single bond dissociation, but return to the potential well region, 
possibly several times, before ultimately dissociating. 
\end{abstract}

\pacs{05.45.-a, 31.15.-p, 34.10.+x, 36.20.-r, 62.25.-g}

\maketitle 


\section{Introduction}
\label{intro}

There has been much recent interest in the rapidly developing field of `mechanochemistry', 
where applied force (e.g., tensile stress) is employed to alter
absolute rates and/or
product ratios of chemical reactions \cite{Kruger03,Huang10,Huang11,Konda11,Wiggins12}.
A fundamental understanding of the intramolecular dynamics and reaction kinetics of
molecules subject to a tensile force 
\cite{Dudko09,Yohichi10,Dudko11,Yohichi11,Makarov11,Prezhdo09}
is needed to provide a solid 
theoretical foundation for mechanochemistry, as well as 
for theories of 
material failure under stress \cite{Kausch87,Crist95,Buehler10},
polymer rupture 
\cite{Kausch87,Crist95,Termonia85,Termonia86,Garnier00,Saitta00,Saitta01,Maroja01,Rohrig01,Suzuki10}
adhesion \cite{Gersappe99}, friction \cite{Fillipov04},
and biological applications of dynamical force microscopy
\cite{Evans01,Williams03,Harris04,Tinoco04,Pereverzev06,Hyeon07,Kumar10}.

Much previous work on the dynamical consequences of the application
of tensile stress has focussed on the investigation of 
fragmentation kinetics of linear chains. 
Studies of energy transfer and equipartition in single chains of coupled anharmonic
  oscillators have played an essential role in the development
  of nonlinear dynamics, beginning with the seminal work of Fermi,
    Pasta and Ulam \cite{Fermi55,Ford92} (see, for example, refs
  \onlinecite{Schranz86a,Schranz86b,Schranz86c,Nyden91,Sumpter92,Sumpter94a,Bolton95,Okabe98,Okabe99,Reigada00,Reigada01,Okabe04,Lepri05}).

The dissociation of a 1-D chain subject to constant tensile force is a problem in
  unimolecular kinetics, and a
  fundamental issue in unimolecular kinetics concerns the applicability of
      statistical approaches such as RRKM \cite{Robinson72,Forst73,Gilbert90,Baer96,Hase98} or
      transition state theory \cite{Wigner38,Keck67,Pechukas81,Truhlar83,Anderson95,Truhlar96}.
  Previous theoretical work has suggested that
    dissociation of atomic chains under stress is not amenable to simple statistical
  approaches
   \cite{Melker79,Melker80a,Melker80b,Melker81,Welland92,Oliveira94,Bolton95,Sebastian99,Puthur02,Sain06}.
  Early trajectory simulations on the dynamics of
  Morse chains \cite{Melker79,Melker80a,Melker80b,Melker81}
  showed that simple bond stretching or force criteria
  for bond rupture were inadequate, in that apparently broken bonds were observed
  to reform (\emph{bond healing}).  
  Subsequent simulations of the fragmentation of
  1-D Lennard-Jones (LJ) chains at constant strain with inclusion of
  a frictional damping term and a stochastic force modelling
  interaction with a heat bath showed that 
  healing of incipient breaks is highly efficient \cite{Oliveira94} 
  (see also refs \onlinecite{Ghosh10,Paturej11}).
  Nonexponential decay, failure of RRKM theory, and extensive
  transition state recrossing effects were found by
  Bolton, Nordholm and Schranz in their studies of the dissociation of 1-D Morse
  chains ($N=2-20$) under stress \cite{Bolton95}.
 
  Standard harmonic classical TST has been
  applied to the dissociation of a 1-D Morse chain \cite{Sebastian99,Puthur02}, with
  the transition state   for dissociation of a given bond located at the  maximum of
  the effective potential (see below).  The {\em harmonic} canonical TST rate
  constant did not agree with
  molecular dynamics calculations, but effects of anharmonicity
\cite{Hase98,Bolton95}
  on the predictions of TST were
  not systematically investigated.
  Both RRKM (fully anharmonic, Monte Carlo) and
    RRK (harmonic appproximation) theory were applied to predict bond dissociation rate
    constants as a function of energy and tensile force 
    for Morse chains under tensile stress \cite{Stember07}.
    For chains with $N \geq 3$ atoms 
    a hybrid statistical theory was used involving a harmonic approximation for motion at the
    transition state for bond dissociation \cite{Stember07}.  More recent work has 
    examined isomerization dynamics for a Morse chain under constant strain subject to
    periodic boundary conditions \cite{Stember11}.

Transition state theory, which has long been a cornerstone
of the theory of chemical reaction rates 
\cite{Wigner38,Keck67,Pechukas81,Truhlar83,Anderson95,Truhlar96},
has been the subject of renewed interest in recent years 
\cite{Wiggins90,wwju,ujpyw,WaalkensBurbanksWiggins04,WaalkensWiggins04,WaalkensBurbanksWigginsb04,
WaalkensBurbanksWiggins05,WaalkensBurbanksWiggins05c,SchubertWaalkensWiggins06,WaalkensSchubertWiggins08,
MacKay90,Komatsuzaki00,Komatsuzaki02,Wiesenfeld03,Wiesenfeld04,Wiesenfeld04a,Komatsuzaki05,Jaffe05,Wiesenfeld05,Gabern05,Gabern06,Shojiguchi08}.
It has been established both theoretically and computationally 
that index one saddles \cite{saddle_footnote1}
of the potential energy surface \cite{Mezey87,Wales03}
give rise to a variety of geometrical structures in {\em phase space},
enabling the realization of Wigner's vision of a transition state theory
constructed in \emph{phase space}
\cite{Wiggins90,wwju,ujpyw,WaalkensBurbanksWiggins04,WaalkensWiggins04,WaalkensBurbanksWigginsb04,
WaalkensBurbanksWiggins05,WaalkensBurbanksWiggins05c,SchubertWaalkensWiggins06,WaalkensSchubertWiggins08,
MacKay90,Komatsuzaki00,Komatsuzaki02,Wiesenfeld03,Wiesenfeld04,Wiesenfeld04a,Komatsuzaki05,Jaffe05,Wiesenfeld05,Gabern05,Gabern06,Shojiguchi08}.

Following these studies, attention has naturally focussed on
phase space structures associated with saddles of index
greater than one, and their possible dynamical significance 
\cite{Ezra09,Haller10,Haller11,Collins11}.
We have investigated phase space structures and their influence on transport
in phase space  associated with  {\em index two saddles} of the potential energy surface
for $n$ degree-of-freedom (DoF) deterministic, time-independent 
Hamiltonian systems \cite{Ezra09,Collins11}.  
We have shown that, for isomerization dynamics in a model 
$n=2$ DoF potential, it is possible to 
distinguish between `concerted' and `stepwise' (or sequential)
isomerization trajectories in a dynamically significant way using
a normal form Hamiltonian describing the phase space structure in the
vicinity of the index 2 saddle.
The importance of index 2 saddles for the question of dynamical separability of
tight versus roaming mechanisms has recently been established \cite{Harding12}.
As discussed in detail in the present paper,
higher index saddles arise naturally in the problem of 
an atomic chain under tensile stress, and we are able to apply the methods and insights
from our previous studies to this problem.

  In the present work normal form theory together with 
  classical trajectory simulations are used to
  investigate the fragmentation kinetics and
  phase space structure of short tethered atomic chains under constant tensile stress.
  Following previous work, we model the interatomic interactions using 
  Morse potentials \cite{Stember07}. 
  Most of our work concerns a `chain' with $n=2$ particles.
  Our focus is on the relation between the phase space structure in the vicinity
  of the index 2 saddle in this system, as described by a  normal form Hamiltonian,
  and the fate of trajectories passing through a suitably defined dividing surface.
  The competition between single and double bond breaking is of particular interest, as
  is the phenomenon of bond `healing' \cite{Oliveira94,Ghosh10,Paturej11}.
  
  Our phase space approach to 
  the problem of single versus double bond breakage in the Morse chain under tensile
  stress is related to the work of Chesnavich on collision induced dissociation 
  \cite{Andrews84,Grice87};
  the results obtained in the present paper
  serve to confirm and extend Chesnavich's insights concerning the nature 
  of the boundaries between phase space
  regions associated with single and double bond breakage.

The structure of this paper is as follows.
In Sec.\ \ref{morsechain} we describe the Hamiltonian for
a tethered Morse oscillator chain under tension.
We analyze the location and types (indices) of equilibria for
the general $n$-particle chain, and provide numerical results for the $n=2$ case.
Specifically, we note the presence of an index 2 saddle in the 2-atom chain
under tensile stress.
In Section \ref{normformsec} we 
discuss phase space structure and dynamics in a neighborhood of a saddle-saddle equilibrium point.
Based on previous work \cite{Ezra09,Collins11}, we define a phase space 
dividing surface (DS) that enables us to  
classify trajectories passing through a neighborhood of the 
saddle point using the values of the integrals associated with the normal form.  We 
also generalize our definition of the dividing surface and define an
\emph{extended dividing surface} (EDS); the EDS can be used to sample and classify trajectories
that pass through a phase space neighborhood of the index 2 saddle at total energies
less or greater than that of the saddle.
In Section \ref{clastrajsec} we describe our classical trajectory studies of
single versus double bond breakage for the $n=2$ chain under tension.  
Initial conditions for trajectories are obtained by sampling the EDS at constant energy, 
and we sample trajectories at fixed energies both above and below the energy of the saddle.  
The fate of trajectories (single versus double bond breakage) is 
explored as a function of the location of the initial condition on the EDS.
We find that, at the boundary between regions of the EDS 
associated with trajectories exhibiting breakage of one or the other bond, there is
either (i) a region coresponding to double bond breakage,
or (ii) a trajectory that is trapped in the vicinity of the potential minimum.
The existence of a layer of doubly dissociative trajectories corresponds to the `coating'
phenomenon noted by Chesnavich in his work on collision-induced dissociation \cite{Andrews84,Grice87}.
Our trajectory results also enable us to identify trajectories that exhibit bond healing; 
such trajectories pass beyond the nominal (index 1) transition state for 
single bond dissociation, but return to the well region possibly several times before
ultimately dissociating. 
Section \ref{conclusec} concludes.
The procedure for sampling the DS and EDS used in the computations reported here
is described in detail in Appendix \ref{appendixsampling}.

\newpage

\section{Morse oscillator chain under tension: Hamiltonian and equilibria}
\label{morsechain}

In this section we  describe the Hamiltonian used to model 
a 1D chain of particles under tensile stress. 
After deriving the equations of motion, we determine the location
of equilibrium points for chains
composed of different numbers of particles. 
We first review the single particle `chain' and then give the general 
form of the Hamiltonian for the $n$-particle  case 
together with analysis of equilibria.

\subsection{One particle Morse chain}

We consider first a one particle `chain'. Although this case is trivial,
the results obtained  will be   
useful when dealing with the general case of $n$-particle chain.

The single degree of freedom (DoF) Hamiltonian for our 1-particle chain is the sum of a kinetic energy 
term and a potential energy described by a Morse oscillator potential \cite{Morse29}
representing a particle of mass $m$ tethered to a wall of infinite mass. 
To the Morse potential we add a linear term in the bond coordinate representing 
a tensile force exerted on the particle, whose magnitude is determined by
a parameter $f$. The one DoF Hamiltonian takes the form:
\begin{equation}
H(x,p;f) = \frac{p^2}{2m}+V_M(x)-f(x-x_e)  \\
\label{1DHam}
\end{equation}
where the Morse potential $V_M(x)$ is 
\begin{equation}
V_M(x)=D_0[1-\exp(-\beta(x-x_e))]^2.
\end{equation}

In the following we will measure energies in units of $D_0$, 
the unperturbed Morse dissociation energy, and length in units of $x_e$, 
the equilibrium bond distance for zero tensile force. 
The Morse parameter $\beta$ will be set equal to $1/x_e$ 
so that in our units its numerical value is $\beta=1$.

Associated Hamiltonian equations of motion (vector field) are:
\begin{subequations}
\label{1Dvecfield}
\begin{align}
\dot{x} & = \frac{\partial H}{\partial p} = \frac{p}{m}
\label{1Dvecfielda} \\
\dot{p} &= -\frac{\partial H}{\partial x} = 
2D_0 \beta \lbrace
\exp\left[-2 \beta (x-x_e)\right] -  \exp\left[ - \beta (x-x_e)\right]\rbrace + f.
\label{1Dvecfieldb}
\end{align}
\end{subequations}

The equilibrium points of this vector field satisfy the following equations
\begin{subequations}
\label{1Dvecfieldfixpoints}
\begin{align}
\dot{x}(x, p) & = 0 
\label{1Dvecfieldfixpointsa} \\
\dot{p}(x, p) & =0 .
\label{1Dvecfieldfixpointsb}
\end{align}
\end{subequations}
Eq.\ \eqref{1Dvecfieldfixpointsa} is always satisfied for $p=0$. 
The number of (real) roots of eq.\ \eqref{1Dvecfieldfixpointsb} 
depends on the value of the $f$ parameter. 
The critical value for this parameter is $f_{crit}={D_0 \beta}/{2}$,
which has the value $f_{crit}=\frac{1}{2}$ in our units. 
For $0<f<f_{crit}$ we have two real roots, for $f=f_{crit}$ we have one real root 
and for $f>f_{crit}$ there is no real root. 
For $0<f<f_{crit}$, the two roots are:
\begin{subequations}
\label{1D2roots}
\begin{align}
x_+ & = -\dfrac{1}{\beta}\, \ln \left( \dfrac{D_0 \beta + \sqrt{D_0^2 \beta^2 - 2D_0 \beta f}}{2D_0 \beta} \right) + x_e
\label{1D2rootsa}\\
x_- &= -\dfrac{1}{\beta} \, \ln \left( \dfrac{D_0 \beta - \sqrt{D_0^2 \beta^2 - 2D_0 \beta f}}{2D_0 \beta} \right) + x_e
\label{1D2rootsb}
\end{align}
\end{subequations}
The equilibrium point $(x_+,0)$ is the equilibrium phase space point
for the Morse oscillator under tensile stress while the $(x_-,0)$ solution 
is a saddle point. (The stability of these two solutions can 
easily be verified by computing the eigenvalues of the matrix associated 
with the linearization of Hamilton's equations about the equilibrium point of interest.)

\begin{figure}[H]
\begin{center}
\includegraphics[scale=1.2]{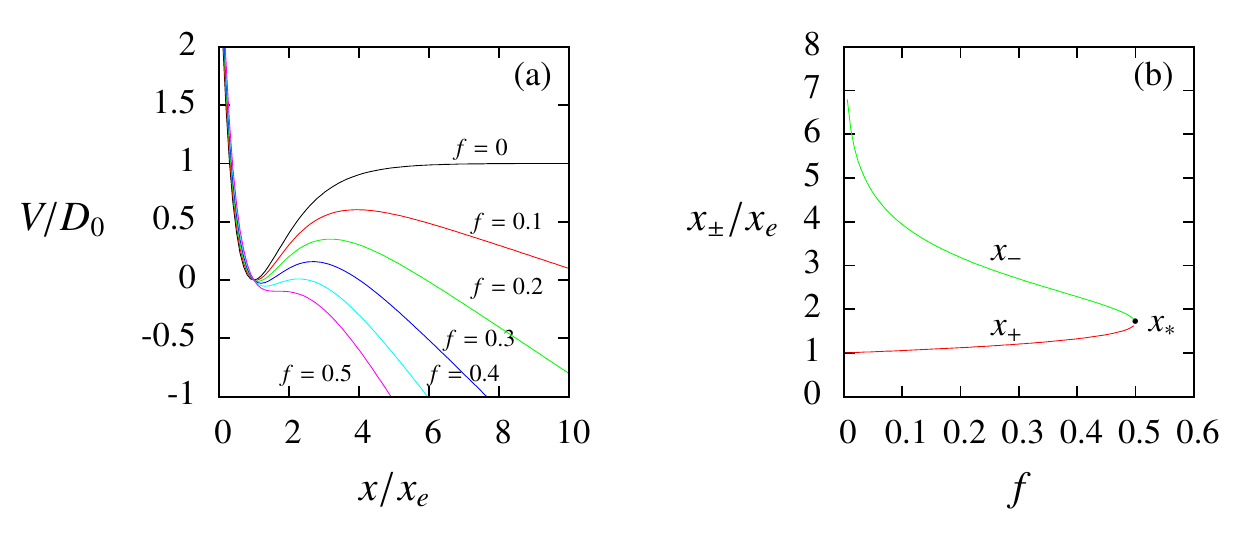}
\caption{(a) Potential function of eq.\ \eqref{1DHam} for different values of 
the force parameter $f$. 
(b) Locations of the equilibrium points of the 1 dof Hamiltonian \eqref{1DHam} as a function of 
the force parameter $f$.}
\label{fig1Dfixpts}
\end{center}
\end{figure}

For $f$ parameter equal to $f_{crit}$ the two equilibrium points $(x_+,0)$ and $(x_-,0)$ 
merge at a single point, which we will denote by $(x_*,0)$, 
in a saddle node bifurcation. For this critical value of $f$ the only real root is
\begin{equation}
x_* = \dfrac{1}{\beta} \ln(2) + x_e
\label{1D1root}
\end{equation}
As the force parameter $f \to 0$, $x_+ \to x_e$, the unperturbed Morse equilibrium distance, 
while $x_- \to  +\infty$. 
Figure \ref{fig1Dfixpts}(a) shows the shape of the 
1-D potential for values of $f$ parameter between $0$ and $f_{crit}$ and 
Figure \ref{fig1Dfixpts}(b) shows the locations of the different equilibrium points 
discussed above as a function of $f$.

\subsection{$n$-particle Morse chain under tension}

 We now consider a chain of $n$ particles tethered to an infinite mass wall. 
 For simplicity we assume all particles have the same mass $m$. 
 
 We use two different sets of phase space coordinates to describe the dynamics. 
 The first set of coordinates are `external' or lab-fixed' variables;
 the configuration space coordinates are denoted  
 $(x_1, x_2, \dots,x_n)$ with conjugate momenta $(p_{x_1},p_{x_2},\dots,p_{x_n})$. 
 The second set of coordinates are `internal' or bond coordinates;
 coordinates are denoted $\mathbf{r} = (r_1, r_2,\dots,r_n)$ with
 conjugate momenta $\mathbf{p_r} = (p_{r_1},p_{r_2},\dots,p_{r_n})$.
 
\begin{figure}[H]
\begin{center}  
\includegraphics[scale=0.25]{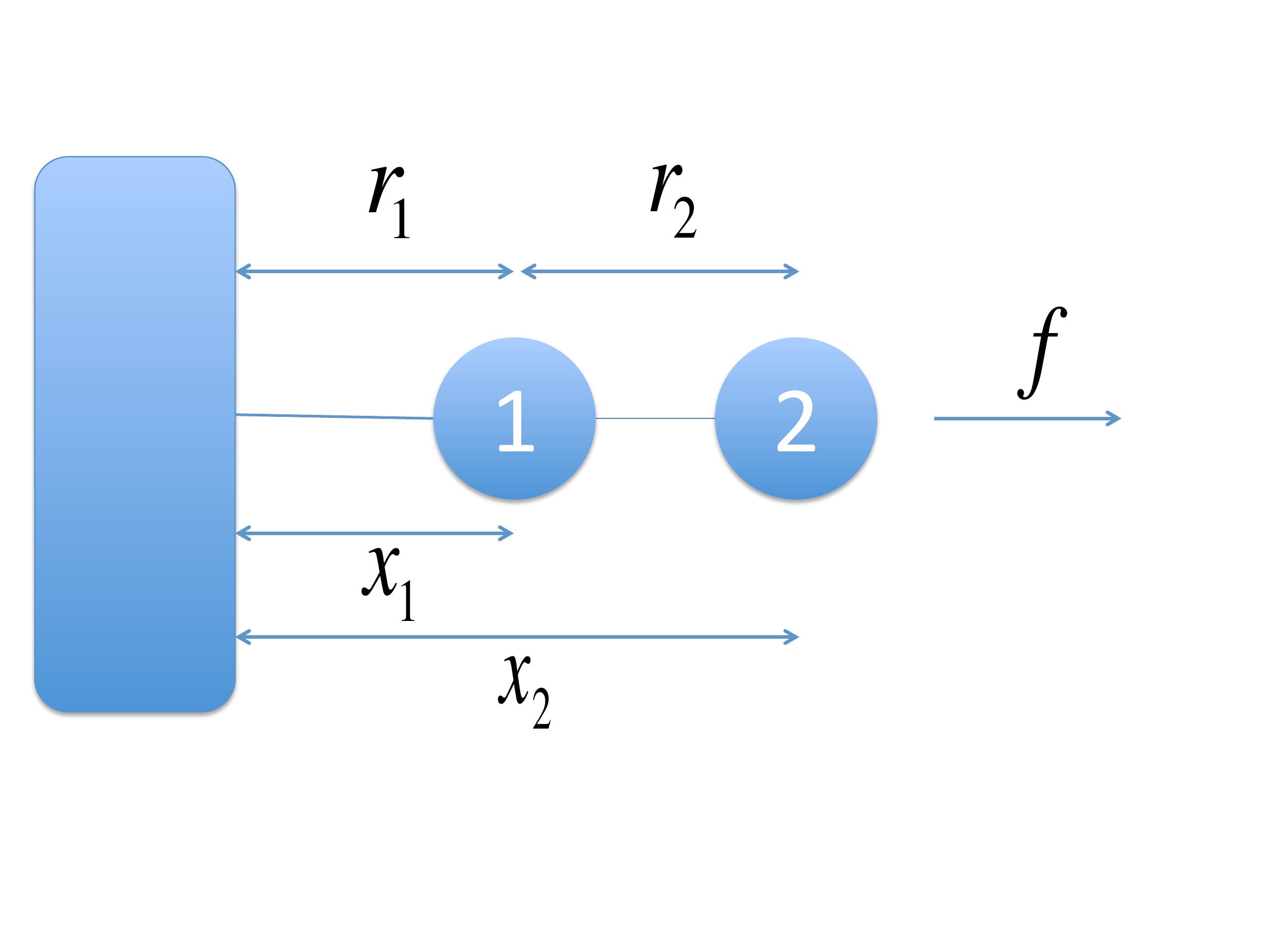} 
\caption{Definition of `external' (space-fixed)  coordinates $(x_1, x_2)$ 
and `internal'  (bond) coordinates $(r_1, r_2)$ 
for the $n=2$ atom Morse chain.}
\label{figcoorddef}
\end{center}
\end{figure}

 Figure \ref{figcoorddef} shows the definitions of these two coordinate systems 
 for the case of a 2-particle chain.  
 The general relation between  coordinates $(x_1, x_2, \dots,x_n)$ 
 and $(r_1, r_2,\dots,r_n)$ is 
\begin{equation}
x_k=\sum_{j=1}^k r_j, \:\: k=1,\dots,n .
\label{relcoordsyst}
\end{equation}

The kinetic energy in external coordinates is a diagonal sum of quadratic terms.
The  potential energy consists of a sum of pairwise Morse interactions 
between adjacent particles together with potential term linear in the  coordinate $x_n$ 
of the last particle
(equivalent to a constant force applied to this particle).  The Hamiltonian is therefore
\begin{equation}
H(x_1,\dots,x_n,p_{x_1},\dots,p_{x_n};f)=\sum_{i=1}^n \frac{p_{x_i}^2}{2m} + V(x_1,\dots,x_n;f)
\label{nDHamextcoord}
\end{equation}
with potential term 
\begin{subequations}
\label{nDpot}
\begin{align}
V(x_1,\dots,x_n;f) & = \sum_{i=1}^n V_M(r_i) - f(x_n-nx_e)
\label{nDpota} \\
& = \sum_{i=1}^n \left[V_M(r_i)-f(r_i-x_e)\right]
\label{nDpotb}
\end{align}
\end{subequations}
In terms of bond coordinates, the Hamiltonian is (cf.\ ref.\ \onlinecite{Stember07})
\begin{equation}
H(\mathbf r,\mathbf{p_r};f)=\frac{p_{r_1}^2}{2m} 
+ \sum_{i=2}^n \frac{p_{r_i}^2}{m} - \sum_{k=1}^{n-1}\frac{p_{r_k}p_{r_{k+1}}}{m} + V(\mathbf r;f).
\label{nDHamintcoord}
\end{equation}

The vector field associated with Hamiltonian \eqref{nDHamintcoord} is:
\begin{subequations} 
\label{nDvecfield}
\begin{align}
\dot{r}_1 & = \frac{\partial H}{\partial p_{r_1}} = \frac{p_{r_1}}{m}-\frac{p_{r_2}}{m}
\label{2Dvecfielda}\\
\dot{r}_i & = \frac{\partial H}{\partial p_{r_i}} = \frac{2p_{r_i}}{m}-\frac{p_{r_{i-1}}}{m}-\frac{p_{r_{i+1}}}{m}, \: i=2,\dots,n-1
\label{2Dvecfieldb}\\
\dot{r}_n & = \frac{\partial H}{\partial p_{r_n}} = \frac{2p_{r_n}}{m}-\frac{p_{r_{n-1}}}{m} 
\label{2Dvecfieldc}\\
\dot{p}_{r_i} & = -\frac{\partial H}{\partial r_i} = 
2D_0 \beta \left[\exp\lbrace-2 \beta (r_i-x_e)\rbrace 
-  \exp\lbrace - \beta (r_i-x_e)\rbrace\right] + f, \: i=1,\dots,n .
\label{2Dvecfieldd}
\end{align}
\end{subequations}
The equilibrium points are found by setting the time derivatives in eq.\
\eqref{nDvecfield} to zero. 
The solutions are in fact easily found in terms of equilibrium points for the $n=1$ DoF system. 
The time derivatives \eqref{2Dvecfielda}, \eqref{2Dvecfieldb} and \eqref{2Dvecfieldc}
are zero only for $p_{r_i}=0$. 
The solutions obtained by setting the time derivatives of the 
momenta to zero depend on the value of $f$ parameter.  
For $0<f<f_{crit}$ we have two real roots for each $r_i$ which are related 
with the one DoF roots by $r_{i\pm}=x_{\pm}$, so that the number of equilibria for 
an $n$-particle chain is $2^n$. The stability of these equilibria is determined by 
the stability of the one DoF equilibrium points, so that we obtain a $c$-centre-$s$-saddle 
when the $x_+$ root occurs $c$ times and the $x_-$ root $s$ times with $n=c+s$. 
The organisation of the different saddle indices follows a
Pascal's triangle structure as shown in table \ref{tablePascaltrig}.

\begin{table}
\begin{tabular}{|c|c|c|c|c|c|c|}
\hline
n\textbackslash k & 0 & 1 & 2 & 3 & 4 & \dots \\
\hline
0                          & - &    &    &    &    &           \\
\hline
1                          & 1 & 1 &    &    &    &           \\
\hline
2                          & 1 & 2 & 1 &    &    &           \\
\hline
3                          & 1 & 3 & 3 & 1 &    &           \\
\hline
4                          & 1 & 4 & 6 & 4 & 1 &           \\
\hline
$\vdots$              & $\vdots$ & $\vdots$ & $\vdots$ & $\vdots$ & $\vdots$ & $\ddots$          \\
\hline
\end{tabular}
\caption{Pascal's triangle structure of the indices of the different 
saddles ($k$) depending on the number of particles in the chain ($n$). 
Index 0 means a stable centre (well).}
\label{tablePascaltrig}
\end{table}

\begin{figure}[H]
\begin{center} 
\includegraphics[scale=1.2]{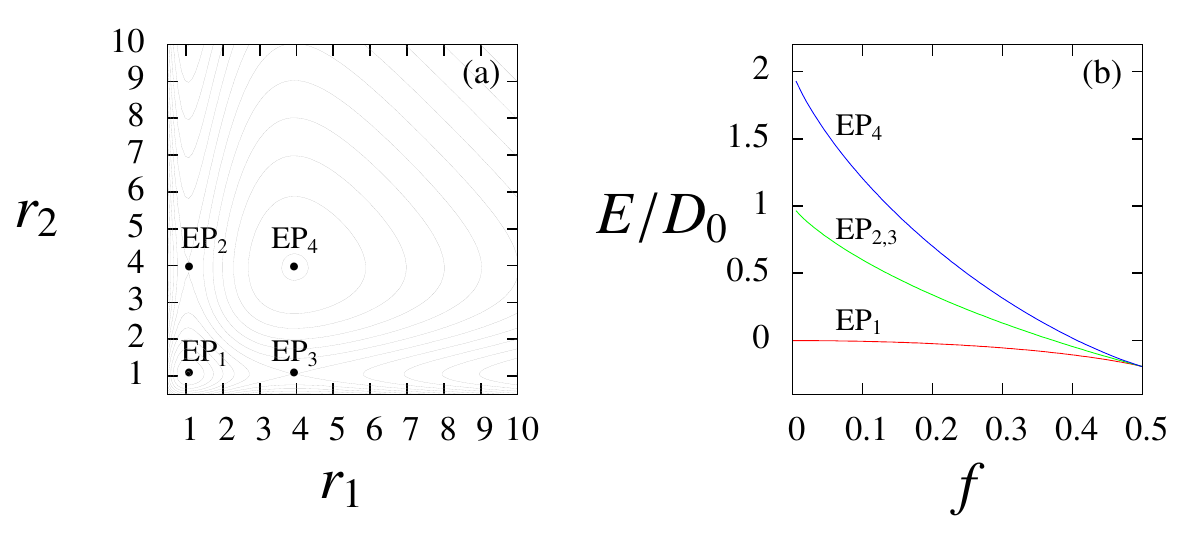}
\caption{(a) Contour plot of the potential energy surface 
for the $n=2$ Morse chain with $f=0.1$. 
(b) Energies of the different equilibria as a function of the force parameter $f$.}
\label{fig2Dfixpts}
\end{center}
\end{figure}

In the rest of this paper we focus on the $n=2$-particle chain. 
For this case we have four equilibrium points for $0<f<f_{crit}$ and 
only one for $f=f_{crit}$. The situation is summarized as follows:
\begin{itemize}
\item $0 < f < f_{crit}$: 
$\text{EP}_1 \equiv (r_{1+},r_{2+},0,0)$, 
$\text{EP}_2 \equiv (r_{1+},r_{2-},0,0)$, 
$\text{EP}_3 \equiv (r_{1-},r_{2+},0,0)$, 
$\text{EP}_4$$\equiv (r_{1-},r_{2-},0,0)$.

\item $f=f_{crit}$: $\text{EP}_{*} \equiv (r_{1*},r_{2*},0,0)$.

\end{itemize}
EP$_1$ is a center, EP$_{2, 3}$ are index 1 saddles and EP$_4$ is an index 2 saddle. 
Figure \ref{fig2Dfixpts}(a) shows contours of the potential function in 
$(r_1, r_2)$ space as well as the location of the equilibria 
while Figure \ref{fig2Dfixpts}(b) shows the evolution of the energy of these equilibria points 
as a function of the force parameter $f$.

\newpage

\section{Phase space structure and dynamics in a neighborhood of the index 2 saddle}
\label{normformsec}

In the present work we focus on the role of the index-2 saddle EP$_4$ 
in the bond dissociation kinetics of the 2-particle chain. 
We study fragmentation dynamics at energies close to the energy of this equilibrium point. 
Our aim here is to investigate the competition between single 
and double bond breakage for trajectories that enter the phase space neighborhood of the
index 2 saddle.
We will study trajectories initiated on a neighbourhood of EP$_4$ and investigate the 
dynamical role of the index 2 saddle on the fragmentation of the chain. 

In order to carry out this program we construct a normal form 
Hamiltonian which provides an integrable approximation to the 
dynamics associated with the original Hamiltonian 
in a neighbourhood of the saddle point EP$_4$. In this section we first briefly 
describe the construction of a normal form Hamiltonian in a neighbourhood of a saddle point;
we then show how this Hamiltonian can be used to describe the 
dynamics in this neighbourhood and different phase space objects such as 
normally hyperbolic invariant manifolds (NHIMs), dividing surfaces 
and extended dividing surfaces (EDS).

\subsection{The normal form}

Normal form theory  has been mainly applied in reaction dynamics involving 
equilibrium points of saddle$\times$center$\times \dots \times$center stability type 
(that is, index 1 saddles). 
For this kind of saddle the normal form Hamiltonian permits the construction 
of certain phase space structures which are of
central importance for the reaction dynamics (for a review and further references, see ref.\
\onlinecite{WaalkensSchubertWiggins08}).

However normal form theory is also useful for describing dynamics in the vicinity of 
equilibria of different stability type such as  EP$_4$ 
in the present problem, which is of saddle$\times$saddle stability 
type.   Recent work has begun to investigate
reaction dynamics mediated by higher index saddles \cite{Ezra09,Haller10,Haller11,Collins11}.

The construction of the normal form coordinate set is carried out 
by Poincar\'{e}-Birkhoff normal form theory.
The details of this theory are by now well-known and have been presented in detail 
in a number of reviews and books (see, for example, 
refs \onlinecite{Wiggins94,WaalkensSchubertWiggins08}). 
Normal form  theory provides an algorithmic procedure for finding a non-linear 
symplectic change of variables,
\begin{equation}
(\mathbf{q},\mathbf{p}) = T(\mathbf{x},\mathbf{p_x}),
\label{changevar1}
\end{equation}
which will transform a given Hamiltonian into a new, simpler,  Hamiltonian,
\begin{equation}
H_{NF}(\mathbf{q},\mathbf{p}) = H( T^{-1}(\mathbf{q},\mathbf{p})) = H(\mathbf{x},\mathbf{p_x}).
\label{changevar2}
\end{equation}
The form of the resulting Hamiltonian $H_{NF}$ is constrained by 
imposing conditions such as the requirement that $H_{NF}$ 
should Poisson commute with a certain Hamiltonian $H_0$; in such a case 
we say that the resulting Hamiltonian is in normal form with respect to this Hamiltonian $H_0$. 
The Hamiltonian $H_0$ is often taken to have the simplest possible form, namely, 
a Hamiltonian describing $n$ uncoupled harmonic oscillators. 

The normal form Hamiltonian will describe in a neighbourhood $\mathcal{L}$ of
the equilibrium point an integrable system which decouples the dynamics 
into ``reaction coordinates'' and ``bath modes''.

In the general case of a $n$ DoF Hamiltonian system, 
the matrix associated with the linearization of Hamilton's equations about the equilibrium point  
has $k$ pairs of real eigenvalues of equal magnitude, but opposite 
in sign $(\pm \lambda_i)$ and $n-k$ pairs of 
complex conjugate purely imaginary eigenvalues $(\pm\omega_i)$. 
We will assume that a non-resonance condition holds between the eigenvalues, 
i.e., the eigenvalues 
$(\omega_{k+1}, \dots, \omega_{n})$ satisfy the relation 
$c_{k+1} \omega_{k+1} + \dots + c_n \omega_{n} \neq 0$ for 
any vector of integers $(c_{k+1},\dots,c_n)$. 
Under these conditions the normalisation procedure transforms the
original Hamiltonian to an even order polynomial in the normal form coordinates 
$(q_1,\dots,q_n,p_1,\dots,p_n)$. 

The normal form Hamiltonian thus obtained, $H_{NF}$, describes an integrable system 
which approximates the dynamics of the original Hamiltonian $H$ in a 
neighbourhood of the equilibrium point. 
As the resulting Hamiltonian is integrable, we can find $n$ integrals 
of motion and express the normal form Hamiltonian explicitly in terms of these integrals:
\begin{subequations}
\label{Kham}
\begin{align}
H_{NF}(\mathbf{q},\mathbf{p}) & = K(I_1,\dots,I_n) \\
& = \lambda_1 I_1 + \dots + \lambda_{k} I_k + \omega_{k+1} I_{k+1} + \dots + \omega_{n} I_{n} + 
\text{{\sc hot}},
\end{align}
\end{subequations}
where the higher order terms ({\sc hot}) are at least of order two in the integrals $(I_1,\dots,I_n)$.
 The expressions of these actions in terms of the normal form coordinates 
 $(q_1,\dots,q_n,p_1,\dots,p_n)$ are: \begin{subequations} 
\label{actions}
\begin{align}
I_i & = q_i p_i, \:\: i=1,\dots,k 
\label{actionsa}\\
I_j & = \frac{1}{2}(q_j^2+p_j^2), \:\: j=k+1,\dots,n.
\label{actionsb}
\end{align}
\end{subequations} 
The vector field associated with the normal form Hamiltonian $H_{NF}$ is 
\begin{subequations} 
\label{nfvectorfield}
\begin{align}
\dot{q}_i &= \frac{\partial H_{NF}}{\partial p_{i}} = \frac{\partial K}{\partial I_i} \frac{\partial I_i}{\partial p_i} = \Lambda_i q_i
\label{nfvectorfielda}\\
\dot{p}_i & = -\frac{\partial H_{NF}}{\partial q_{i}} = -\frac{\partial K}{\partial I_i} \frac{\partial I_i}{\partial q_i} = -\Lambda_i p_i, \:\: i=1,\dots,k
\label{nfvectorfieldb}\\
\dot{q}_j &= \frac{\partial H_{NF}}{\partial p_{j}} = \frac{\partial K}{\partial I_j} \frac{\partial I_j}{\partial p_j} = \Omega_j p_j
\label{nfvectorfieldc}\\
\dot{p}_j &= -\frac{\partial H_{NF}}{\partial q_{j}} = -\frac{\partial K}{\partial I_j} \frac{\partial I_j}{\partial q_j} = -\Omega_j q_j, \:\: j=k+1,\dots,n
\label{nfvectorfieldd}
\end{align}
\end{subequations}
where we have defined the frequencies
\begin{subequations} 
\label{freq}
\begin{align}
\Lambda_i & = \frac{\partial K}{\partial I_i}, \:\: i=1,\dots,k
\label{freqa}\\
\Omega_j & = \frac{\partial K}{\partial I_j}, \:\: j = k+1,\dots,n.
\label{freqb}
\end{align}
\end{subequations}
We now introduce a new set of canonical coordinates, $(Q_i,P_i)$, for the saddle planes. 
These coordinates are useful for describing the dynamics in each saddle plane
and are obtained by a canonical transformation of the NF variables $(q_i,p_i)$ $i=1,\dots,k$
\begin{subequations} 
\label{barcoord}
\begin{align}
Q_i & = \frac{1}{\sqrt{2}} (q_i - p_i)
\label{barcoorda}\\
P_i & = \frac{1}{\sqrt{2}} (q_i + p_i), \:\: i=1,\dots,k .
\label{barcoordb}
\end{align}
\end{subequations}
In terms of these coordinates the action variables are
\begin{equation}
I_i = \frac{1}{2} (P_i^2 - Q_i^2), \:\: i=1,\dots,k
\label{actionsbarcoorda}
\end{equation}
while the vector field for the saddle modes transforms to:
\begin{subequations} 
\label{nfvectorfieldbar}
\begin{align}
\dot{Q}_i & = \frac{\partial H_{NF}}{\partial P_{i}} = \frac{\partial K}{\partial I_i} \frac{\partial I_i}{\partial P_i} = \Lambda_i P_i
\label{nfvectorfieldbara}\\
\dot{P}_i & = -\frac{\partial H_{NF}}{\partial Q_{i}} = -\frac{\partial K}{\partial I_i} \frac{\partial I_i}{\partial Q_i} = \Lambda_i Q_i, \:\: i=1,\dots,k
\label{nfvectorfieldbarb}
\end{align}
\end{subequations}
and the vector field for the bath modes is unchanged.

\subsection{Phase space structures in normal form coordinates}

\subsubsection{Crossing and non-crossing trajectories}

The normalization algorithm provides us with an integrable system. 
The dynamics separates into independent motions in the
position-momentum planes for the saddle modes 
and the bath modes, so that the full dynamics is the cartesian product of the motion
in these planes. 

There are $k$ saddle (`reactive') planes spanned by coordinates $(q_i,p_i)$ or 
$(Q_i,P_i)$,  $i=1,\ldots,k$. The phase space for each saddle mode 
is foliated by manifolds specified by the value of the integral of motion $I_i$, $i=1,\dots,k$. 
For each value of $I_i$, the trajectory curves (solutions of 
Hamilton's equations) are simply the two branches of the hyperbola, 
\begin{equation}
q_ip_i=\frac{1}{2}(P_i^2-Q_i^2)=I_i .
\end{equation}
The dynamics in the remaining 
$n-k$ bath modes consists of uniform rotation in angle $\theta_j$ conjugate to the
conserved action $I_j$ in the respective $(q_j, p_j)$ planes with $j=k+1,\dots,n$.
The total energy $E$ of the normalized system is a function of the action
variables only:
\begin{equation}
K(I_1,\dots,I_n) = E.
\label{HequalE}
\end{equation}

The $k$ saddle integrals $I_i$, $i=1,\dots,k$, 
can either be positive or negative, leading to two different types of branches of 
the hyperbola in each saddle plane depending on the sign of the integral. 
For $I_i>0$ the two branches of the hyperbola form what we call \emph{crossing trajectories} 
whereas those branches for which $I_i<0$ are called \emph{non-crossing trajectories}. 
Considering constant action hyperbolae in the $(Q_i,P_i)$ plane, crossing trajectories are 
those for which the sign of $Q_i$ changes along the branch whereas non-crossing trajectories
are those for which the sign of $Q_i$ remains the same. For crossing trajectories, 
we can distinguish trajectories for which the sign of $Q_i$ changes from negative to positive
from trajectories for 
which the sign of $Q_i$ changes from positive to negative.

\begin{figure}[H]
\begin{center}  
\includegraphics[scale=1.0]{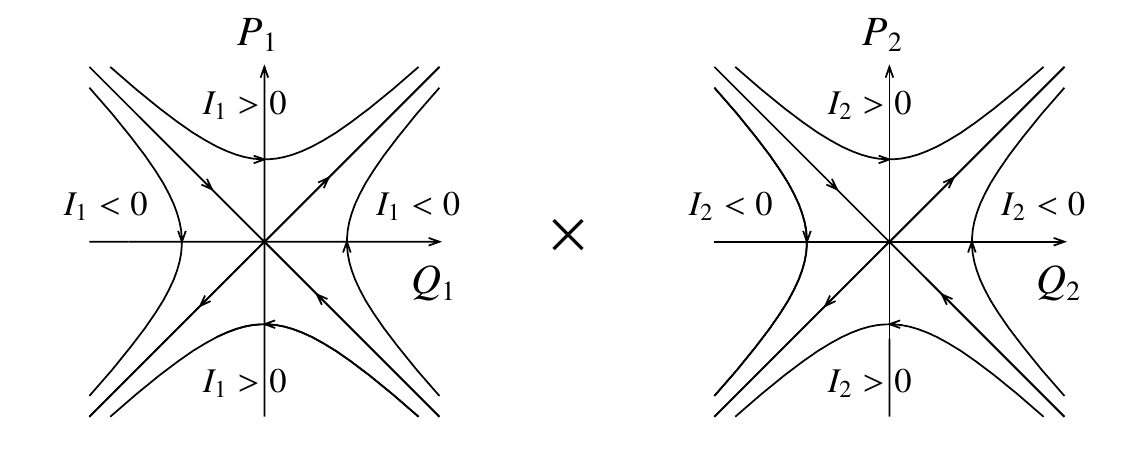}
\caption{Schematic representation of the dynamics in 
the two saddle planes in physical normal form coordinates.}
\label{fsaddlesplanes}
\end{center}
\end{figure}

As discussed in ref.\ \onlinecite{Ezra09}, 
we can introduce a symbolic description of trajectories in 
the neighbourhood of the origin in the saddle plane. 
A trajectory is labelled by 2 symbols, ($f; i$), 
where $i=\pm$, $f=\pm$. 
Here $i$ refers to the initial sign of $Q_j$ as it enters the neighbourhood of 
the origin and $f$ refers to the final sign of $Q_j$ as it leaves the neighbourhood. 
In the case of multiple saddle mode planes, we can extend this symbolic 
description to $2k$ indices by taking the cartesian product of the $k$ saddle planes and
labelling a trajectory by ($f_1 \: \dots \: f_k;i_1\: \dots \: i_k$). 
For example, with two saddle modes, a trajectory labelled ($+- ;- -$) 
crosses in the first plane and does not cross 
in the second plane, the sign of $Q_2$ remaining negative. 
Figure \ref{fsaddlesplanes} shows a schematic representation of the 
different hyperbola in the two saddle planes.

\subsubsection{Dividing surfaces, extended dividing surfaces and NHIMs}

The notion of a dividing surface originates in the study of chemical reaction dynamics. 
In this context one is interested in defining a  
surface in phase space separating reactants from products,
through which all 
reactive trajectories must pass,  and which is never encountered by any 
nonreactive trajectories. This surface is conventionally 
referred to as the transition state (TS). Transition state theory (TST) 
has a long history going back to Eyring \cite{Eyring35}, Wigner \cite{Wigner38}
and Keck \cite{Keck67} (variational transition state theory, 
introduced in order to deal with the problem of recrossing \cite{Truhlar83,Anderson95}).

Many chemical applications of TST employ a TS defined in configuration space \cite{Truhlar96}.
For the case of 2 DoF, 
a dynamically based approach to TST was pioneered by Pechukas, Pollak and Child, who
introduced the notion of periodic orbit dividing surface (PODS) \cite{Pechukas81}.

There has been significant 
recent progress in generalizing the dynamically based PODS approach to 
obtain a definition of dividing surfaces in phase space  for $n \geq 3$ DoF.
The phase space approach uses normal form theory 
to construct a normal form Hamiltonian which reproduces the dynamics 
in the neighbourhood of the phase space region of interest \cite{WaalkensSchubertWiggins08}. 
This normal form Hamiltonian is the key object which enables the 
precise mathematical realization of the intuitive idea of a `surface of no return' 
intersecting all reactive trajectories
\cite{Wiggins90,wwju,ujpyw,WaalkensBurbanksWiggins04,WaalkensWiggins04,WaalkensBurbanksWigginsb04,
WaalkensBurbanksWiggins05,WaalkensBurbanksWiggins05c,SchubertWaalkensWiggins06,WaalkensSchubertWiggins08,
MacKay90,Komatsuzaki00,Komatsuzaki02,Wiesenfeld03,Wiesenfeld04,Wiesenfeld04a,Komatsuzaki05,Jaffe05,Wiesenfeld05,Gabern05,Gabern06,Shojiguchi08}.

We now turn to the definition of the dividing surface (DS) and the 
so-called extended dividing surface (EDS) for the present problem in which there 
is a dynamically significant  index 2 saddle point. 
The DS and the EDS are codimension one surfaces within the constant energy surface that 
are transverse to the vector field defined by Hamilton's equations, 
and consist of the set of phase space points for which a suitable 
distance (to be defined) from the equilibrium point is (locally) minimal 
along the trajectory passing through the phase space point.

Whereas the energy at which the DS is defined cannot 
be arbitrary, the EDS is defined for any value of the energy. 
For simplicity we take the zero of energy to be the energy of the 
equilibrium point in the neighbourhood of which we compute the normal form Hamiltonian. 

We define the square of the distance from the origin
\emph{in phase space} to be

\begin{equation}
D \equiv \frac{1}{4} \sum_{i=1}^k \left( Q_i^2 + P_i^2 \right)
\label{defdistance}
\end{equation}

\noindent
where the sum is taken over the saddle DoF.
(Note that a different definition of the distance $D$ was used in ref.\
\onlinecite{Collins11}.  For further discussion of this point, 
see Appendix \ref{appendixsampling}.)
Requiring that points on the DS and the EDS are those for which 
this distance is minimized 
implies that the time derivative of this distance 
along the trajectory passing through the phase point
should vanish:

\begin{equation}
\dot{D} = \sum_{i=1}^k \Lambda_i Q_i  P_i = 0.
\label{defdistance_2}
\end{equation}

\noindent
This relation enables us to 
find those phase space points within a given energy surface 
that satisfy the (local) minimum distance requirement. 

The EDS  is defined by the intersection of the following $2n-1$ dimensional surfaces in the $n$-dimensional phase space:

\begin{eqnarray}
S_1 (Q_1, P_1, \ldots , Q_k, P_k, q_{k+1}, p_{k+1}, \ldots , q_n, p_n) & = &  K(I_1, \ldots , I_n) -E = 0, \nonumber \\
S_2 (Q_1, P_1, \ldots , Q_k, P_k, q_{k+1}, p_{k+1}, \ldots , q_n, p_n) & = &  \sum_{i=1}^k \Lambda_i Q_i  P_i = 0
\label{def:EDS}
\end{eqnarray}

\noindent
The DS is obtained from \eqref{def:EDS} by restricting to points that lie on crossing trajectories.
This condition implies that the actions associated with the saddle 
degrees-of-freedom should be positive. 
On the other hand, points on the EDS are not required to lie on
crossing trajectories so that all phase space points within a 
given energy surface satisfying the minimum distance requirement belong to the EDS. 

\begin{figure}[H]
\begin{center} 
\includegraphics[scale=0.6]{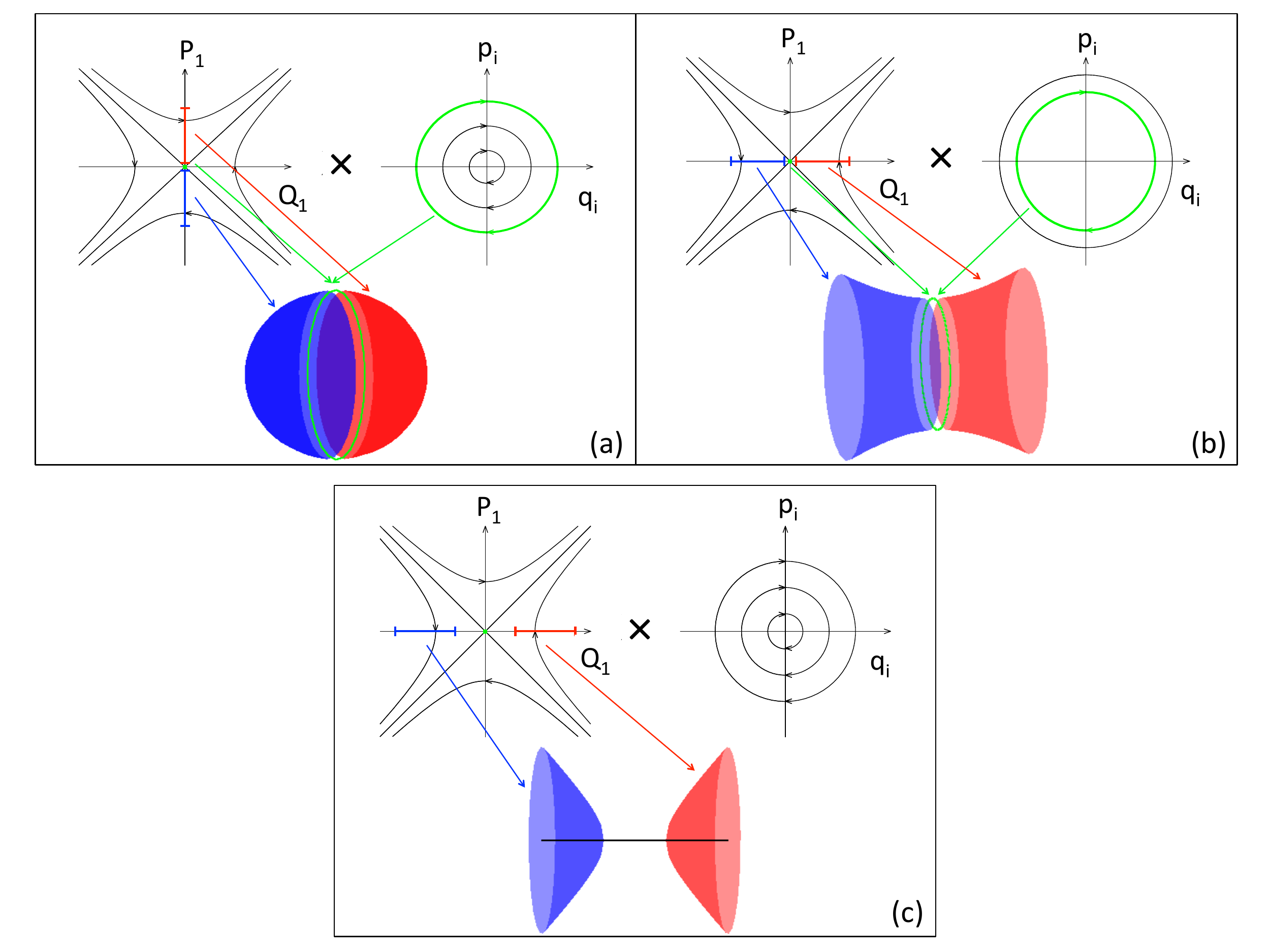}
\caption{}
\label{DSNHIMsphere}
\end{center}
\end{figure}

In the simplest and most familar case, 
a chemical transformation (reaction) involves an equilibrium point of 
saddle$\times$centre$\times \dots \times$centre stability type.
The constant energy dividing surface intersects all reactive trajectories 
and consists of points for which the phase space distance 
from the saddle point is a minimum. 
Clearly, for a trajectory to react, the action associated with
the saddle mode must be 
positive in order for the trajectory to overcome the barrier, 
whereas the actions relative to the bath are always positive or zero;
the energy at which the dividing surface is defined can only be positive. 
The set of points in a given energy surface which minimize 
the distance from the saddle equilibrium point are then 
located precisely on the line $Q_1=0$ 
in the saddle mode plane and in a disk in the bath mode planes. 
When all the energy is in the saddle direction, the saddle action is 
a maximum and the actions relative to the bath modes are all zero
so that $K(I_{1max},0,\dots,0) = E >0$. 
Conversely, when all the energy is distributed amongst the bath modes, 
the action of the saddle is zero; such phase points do 
not belong to the DS and in fact constitute a normally hyperbolic invariant manifold, or  NHIM \cite{Wiggins94}. 
For the case of a 2-DoF system with one saddle mode and one bath mode this NHIM 
is a periodic orbit, the so-called periodic orbit dividing surface (PODS) \cite{Pechukas81}. 
This PODS formally divides the DS into two disjoint pieces: 
one part for which the reactive trajectories cross the DS from reactants to products 
(forward reaction) and one piece for which they cross the DS from products to reactants 
(backward reaction). Again, for a 2-DoF system, each dividing surface 
(forward/backwards) is a disk, 
so that the union of the DS and the NHIM is topologically equivalent 
to a sphere, as  represented in Figure \ref{DSNHIMsphere}(a), 
where the NHIM (PODS) is the equator of this sphere.
The NHIM, 
represented in green, separates the sphere into 
two hemispheres, the forward (red) and backward (blue) hemispheres.

The phase space definition of the DS is essential for a fundamental
understanding of reaction dynamics \cite{Pechukas81,WaalkensSchubertWiggins08}. 
On the other hand, if we are interested in characterising all possible types of  
dynamics in the vicinity of a saddle equilibrium, the DS as defined above does not provide
complete information, as (by definition) it intersects only those trajectories which react. 
In order to define an object which will capture the totality of the dynamics 
in the phase space neighborhood of the equilibrium point we must extend the notion of a 
dividing surface by relaxing the restriction to crossing trajectories
and so include non-crossing trajectories as well. 

This new object is called the \emph{extended dividing surface} (EDS). 
Formally the definition of this surface is a surface within an energy surface consisting
of phase points which (locally) minimize the distance from the saddle point along their
trajectory, without any restriction to crossing trajectories. 
As a consequence, the action associated with the saddle mode 
can now be either positive or negative. 
While the DS was only defined for positive energies, the EDS 
is defined for energies either positive or negative. 
The topology of the EDS depends, however, on whether 
the energy is positive or negative.

Consider the case of an index-1 saddle.  For positive values of the saddle action $I_1$
the EDS is simply the usual DS \cite{Collins11};
that is to say, it is
a sphere with forward and backward hemispheres separated by a NHIM. 
If we allow the saddle action to be negative, for $E>0$
the equation $K(I_{1}<0,I_2,\dots,I_n) = E >0$ must be  satisfied. 
In order to compensate for the loss of energy in the saddle direction, 
the actions of the bath modes have to increase. Topologically the EDS is equivalent to a
hyperboloid as represented schematically in Figure \ref{DSNHIMsphere}(b). 
Notice that in this case the EDS is not compact  as 
the action of the saddle mode can become arbitrarily large and negative,
with a corresponding increase of the bath 
mode actions to conserve energy. Of course this picture is  valid only in the neighbourhood of validity of the normal form Hamiltonian and will 
in general break down outside this neighbourhood.  It is important to realise that the EDS only has meaning within this (bounded) neighbourhood of validity for the normal form.

For negative energies the DS is not a subset of the EDS and this EDS 
is topologically equivalent to a two sheeted hyperboloid represented in Figure \ref{DSNHIMsphere} (c).

For an index 1 saddle in an $n$ DoF system, 
the EDS is of subsidiary importance to the DS for understanding
phase space structure and reaction dynamics.
When we consider the case of higher index saddles, however,
the EDS becomes an object of central importance. 
For this more complex situation the notion of dividing surface itself becomes 
problematic due to the fact that concepts like `reaction pathway' 
are not necessarily well-defined (see ref.\ \onlinecite{Harding12} and refs therein). 
Even if we retain the strict mathematical definition of the DS as the surface 
which intersects all reactive trajectories, the resulting surface 
gives a rather restricted view of the dynamics in the vicinity of the equilibrium point,
whereas the EDS is an  object encoding all information about the dynamics 
in the vicinity of the equilibrium.

We now consider in detail the case of an $n$ DoF system with 
an index 2 saddle. In this case we have two saddle planes and $n-2$ bath mode planes. 
For both positive or negative energies we can always find a combination 
of the different actions which satisfy the equation $K(I_1,I_2,\dots,I_n) = E$ 
with $I_{1,2}$ positive or negative and the bath actions always positive. 
Points on the EDS projected onto the saddle plane $(Q_i, P_i)$ 
can therefore be located on any quadrant of the saddle plane. 

Fixing a point in one of the saddle planes, $(Q_1,P_1)$ say, 
corresponding phase points on the EDS 
must minimize the distance from the equilibrium point.
In Appendix \ref{appendixsampling} we describe our procedure for
locating phase points on the EDS at fixed energy.
Specifically, we show how a parametrization of the hyperbola describing the 
NF dynamics 
in each saddle plane can be used to sample the EDS. Parameters $t_i$, $i=1,\dots,k$, 
describe the position of the phase point on these hyperbolae; a 
useful representation of the EDS therefore utilizes the space 
of these $t_i$ parameters. 

\begin{figure}[H]
\begin{center}
\includegraphics[scale=0.5]{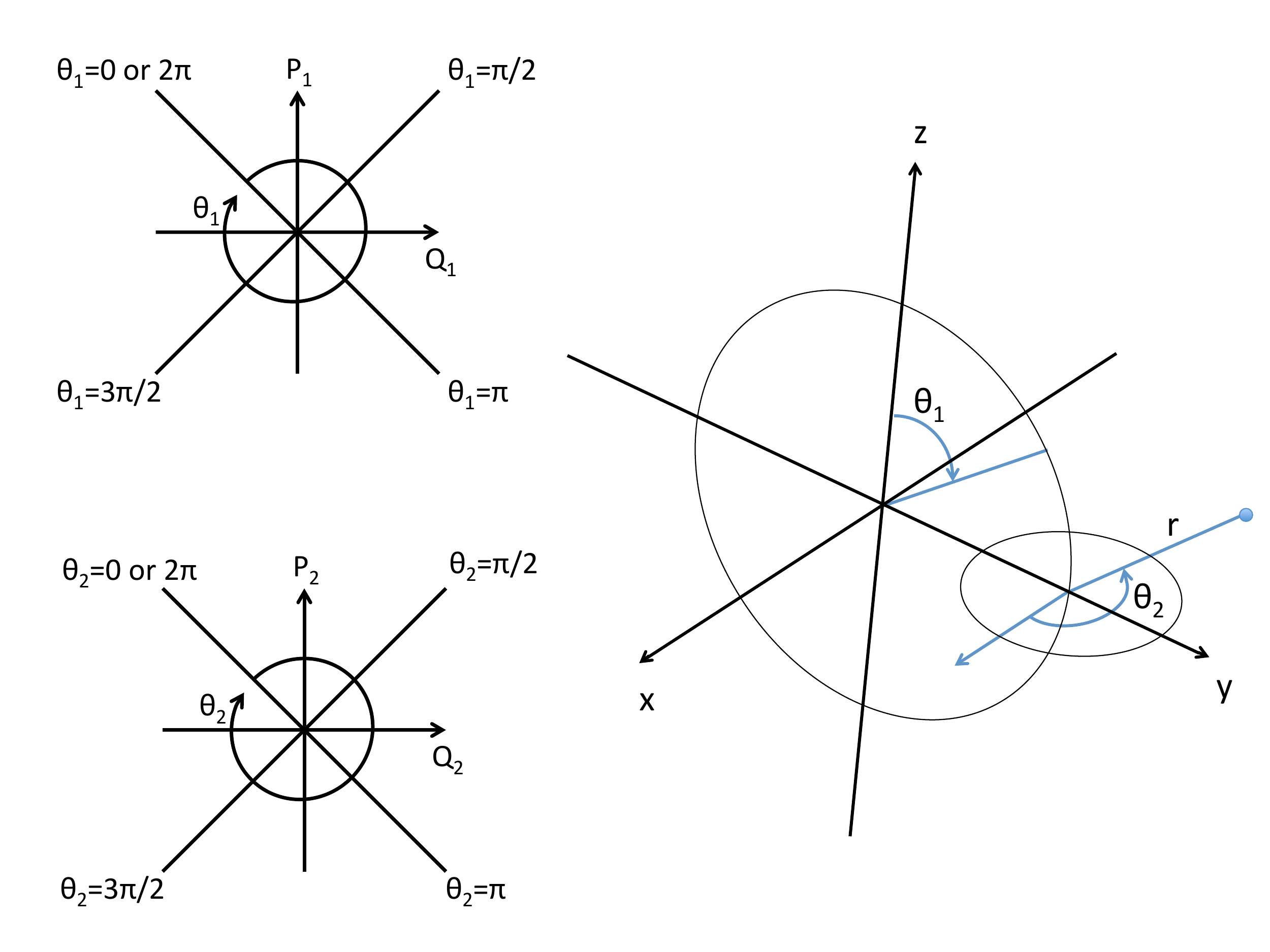}
\caption{Toroidal representation of the extended dividing surface.}
\label{toroidalrep_EDS}
\end{center}
\end{figure}

As discussed in Appendix \ref{appendixsampling}, such a representation is multivalued: 
for every choice of the $t_i$ parameters, there 
will be several phase space points on the EDS corresponding to 
different combinations of signs of the actions $I_i$.
It is useful to have a representation of the EDS
that retains crucial information such as location of a phase point
in the saddle planes and the distance from the equilibrium. 
In a saddle plane, the character of a trajectory depends on which quadrant 
the trajectory intersects the EDS. 
The location of the phase point in the saddle plane can be specified by an angle;
for the case of an index 2 saddle there are two angles. 
We use these two angles and the distance in phase space from the saddle point
to construct a toroidal representation of the EDS. 
Figure \ref{toroidalrep_EDS} shows the definitions of the angles used
and the toroidal representation of the EDS. The torus topology is appropriate 
because phase space points are periodic functions of the angles $t_i$
with period $2 \pi$ .

As phase points composing the EDS can belong to crossing or 
non-crossing trajectories,
the EDS splits into several parts 
(four parts for each saddle plane, $4^k$ parts for an index $k$ saddle). 
In order to label these different parts of the EDS, 
we can use the symbolic description defined for crossing and non-crossing trajectories. 
For the case of $k=2$ saddle modes the 16 parts of the EDS 
are labelled by four indices, ${(f_1 \: f_2;i_1 \: i_2)}$. 
For example, the part of the EDS having trajectories which cross 
in the first saddle plane and which 
do not cross in the second saddle plane (with $Q_2 >0$) 
is denoted EDS$_{(++;-+)}$. 
Notice that, according to our definitions, 
the DS is a subset of the EDS consisting of
the parts with symbol codes ${(++;--)}$, ${(+-;-+)}$, ${(-+;+-)}$ and ${(--;++)}$.

An important question concerns the method  used to sample points on the EDS. 
One method was described in a previous work \cite{Collins11}. 
In the present paper we use another sampling method for the EDS, 
which is described in detail in Appendix \ref{appendixsampling}.

\newpage

\section{Single versus double bond breakage for the $n=2$ chain}
\label{clastrajsec}

In this section we use classical trajectories to investigate single versus double bond breakage 
in the $n=2$ particle chain. By `double bond breakage', we mean fission of both bonds leading 
to the complete breakup of the chain.  Specifically,
we investigate the behavior of trajectories that pass through a neighborhood
of the index 2 saddle.  Similar dynamics was explored in the pioneering
work of Chesnavich on collision induced dissociation reactions  \cite{Andrews84,Grice87}.  

Our problem has two essential parameters.
The first is the magnitude $f$ of the tensile force. 
Varying this parameter changes both the absolute and relative energies 
of the various equlibrium points: the well (EP$_1$), 
the two index one saddles (EP$_{2,3}$), and the index two 
saddle (EP$_4$). In the present work we set $f=0.1$, 
which represents a physically realistic value of 
the tensile stress \cite{Stember07}.
The other parameter of interest is the energy at 
which we study the breakage of the chain. Specifically, we are
interested in the energy dependence of single versus double breakage of the chain 
at energies close to the energy of the index two saddle (i.e., near threshold).

To study these questions, we propagate classical trajectories  
and examine their fate. Points on the EDS (defined as in the previous Section) 
are used as initial conditions for trajectory propagation. 
We examine the dependence of the fate of trajectories on their location on the EDS,
and attempt to understand the distribution over the EDS in terms of 
single breakage of either the first or the second bond and double breakage. 
We also investigate the evolution of  this distribution 
as we change the energy at which the EDS 
is sampled from an energy greater than that of the index two saddle EP$_4$ to 
an energy below. Finally we discuss a possible manifestation
and interpretation of the phenomenon of \emph{bond healing} \cite{Oliveira94,Ghosh10,Paturej11}
in the light of the behavior found in our classical trajectory simulations.

\subsection{General considerations}

To investigate the fate of trajectories at a fixed energy in 
the neighbourhood of the index 2 saddle EP$_4$ for the $n=2$ particle chain, 
we apply normal form theory to construct a Hamiltonian 
which approximates
the dynamics in a neighbourhood of the equlibrium EP$_4$. 
This equilibrium point is of saddle$\times$saddle stability type 
(for $n=2$ the normal form is a 2 DoF Hamiltonian without bath modes). With this NF Hamiltonian in hand, the 
sampling of the EDS is carried out using the procedure described in Appendix \ref{appendixsampling}. 
Using the backward transformation  from normal 
form coordinates to physical coordinates defined in equations 
\eqref{changevar1} and \eqref{changevar2}, we obtain phase space points belonging to the 
EDS in the original physical coordinates. 
We use these points as initial conditions for trajectory propagation, 
and monitor the fate of each trajectory.

Several different types of trajectory are possible.
The first, denoted type 0, are trajectories which remain trapped in the 
well region in the neighbourhood of  equilibrium (minimum) EP$_1$ over the 
full integration time. 
For these trajectories no bonds rupture. 
The second type of trajectories, denoted type 1,  are those which 
exhibit single bond breakage; those trajectories for which bond $r_1$ breaks will be 
denoted type 1$_1$ and those for which bond $r_2$ breaks 
type 1$_2$. Finally, type 2 trajectories are those for which both 
bonds break, corresponding to complete fragmentation of the chain. 

One important question concerns the criteria used to 
decide whether or not a bond has actually broken.
In the work reported here, we propagate trajectories for a long time and look 
at the values of the bond lengths $r_1$ and $r_2$ at the end of the run. By choosing 
thresholds for both $r_1$ and $r_2$ coordinates we can determine  
the trajectory type by comparing the final values of the bond coordinates of the 
propagated trajectory with suitably chosen values. For our simulations, 
we integrate trajectories for 2000 time units (time units 
as defined in Section \ref{morsechain}) and the thresholds in $r_1$ 
and $r_2$ coordinates were chosen to be 30 (units of $r_e$).

\begin{figure}[H]
\begin{center} 
\includegraphics[scale=0.4]{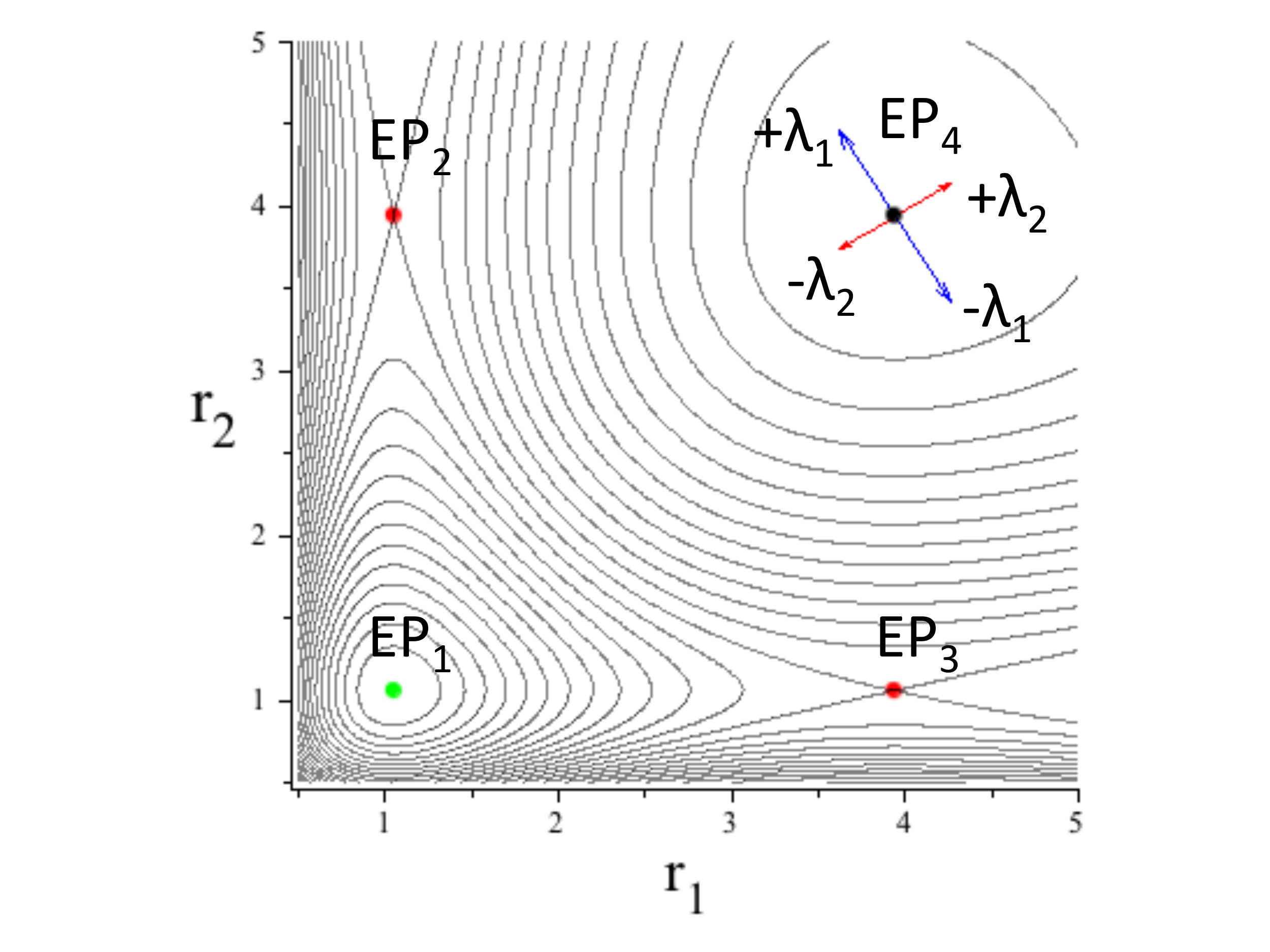}
\caption{Potential energy contours and 
projections of eigenvectors obtained by linearizing 
Hamilton's equations about the equilibrium point EP$_4$.}
\label{poteigenvectors}
\end{center}
\end{figure}

The matrix associated with the linearisation of Hamiltons equations 
about the equilibrium point EP$_4$ has two pairs of real eigenvalues of 
equal magnitude and opposite signs. For the present problem, 
with tensile stress parameter $f=0.1$, 
these eigenvalues are $\pm \lambda_1 = \pm0.4972$ and $\pm \lambda_2 = \pm0.1899$. 
As is shown in Figure \ref{poteigenvectors}, the projections of the 
associated eigenvectors into the $(r_1, r_2)$ plane
indicate reactive directions in the 
two saddle planes given by the normal form Hamiltonian. 
Roughly speaking,  the `direction' associated with the first eigenvalue provides 
in a sense an indicator for which bond will be broken. If we cross 
the saddle in this eigendirection from negative to positive $Q$, we will follow the direction 
given by the eigenvector associated with $+\lambda_1$ and 
consequently we are likely to break bond $r_2$. If we cross in the opposite direction, 
we follow the direction given by the eigenvector associated with $-\lambda_1$ 
and are likely to break bond $r_1$. The directions corresponding to
the eigenvector associated with 
the second eigenvalue indicate whether the trajectory will enter
the well region (in the neighbourhood of EP$_1$) or pass into the region
where we expect double bond breakage to occur. 
This crude approach to the dynamics for trajectories entering the vicinity of the index 2 saddle
does not capture the full complexity of the dynamics; 
in fact there is a
competition between these two directions 
in determining the behaviour of the trajectories.

\subsection{Results}

\subsubsection{Energy above the index 2 saddle}

We consider first the behaviour of trajectories initiated on the EDS 
for energies above the energy of the EP$_4$  saddle point. 

To understand our results, we examine the 
dynamics associated with the second saddle plane in normal form coordinates; 
roughly speaking, motion along the associated eigendirection 
determines whether or not the trajectory 
enters the well region or exits directly into a region of configuration space where 
we might expect double bond breakage to occur. 
The normal form Hamiltonian in this plane has 4 distinct types of dynamics
(combinations of symbols $\pm$):
two correspond to crossing trajectories and two to non-crossing trajectories. 
Non-crossing trajectories remain 
either in the well region ($Q_2 <0$) or outside the well 
region ($Q_2>0$). There are therefore two relevant classes of trajectories. The first class is 
composed of trajectories for which the sign of 
$Q_2$ changes from negative to positive together with non-crossing trajectories 
for which $Q_2 > 0$. The second class is composed of 
trajectories  for which the sign of $Q_2$ 
changes from positive to negative together with non-crossing trajectories for 
which $Q_2 < 0$ remain negative. Using the symbolic classification,
the first class consists of parts of the EDS associated with codes 
${(f_1+;i_1-)}$ and ${(f_1+;i_1+)}$, and the second class 
with codes ${(f_1-;i_1+)}$ and ${(f_1-;i_1-)}$, where $f_1$ and $i_1$ 
are used to designate all the possible symbols for the first saddle plane ($Q_1,P_1$).

Let us focus on the first of these classes, ${(f_1+;i_1-)} \cup {(f_1+;i_1+)}$. 
This class splits into two different subclasses: ${(f_1+;i_1-)}$ and ${(f_1+;i_1+)}$. 
Each of these two subclasses splits in turn into several 
distinct subsubclasses according to the symbols 
associated with motion in the first saddle plane. 
The first subclass is composed of the parts  ${(++;+-)}$, ${(++;--)}$, ${(-+;--)}$ 
and ${(-+;+-)}$, whereas the second subclass is composed of 
the parts ${(++;++)}$, ${(++;-+)}$, ${(-+;-+)}$ and ${(-+;++)}$. 
The sets of points on the EDS associated with symbol codes
${(++;++)}$ and ${(-+;-+)}$ are actually empty. 
This is due to the fact that, when one samples the EDS and tries to 
solve the equation $K(I_1,I_2)=E>0$, there is no solution 
for which the two actions are negative simultaneously. 

\begin{figure}[H]
\begin{center} 
\includegraphics[scale=1.0]{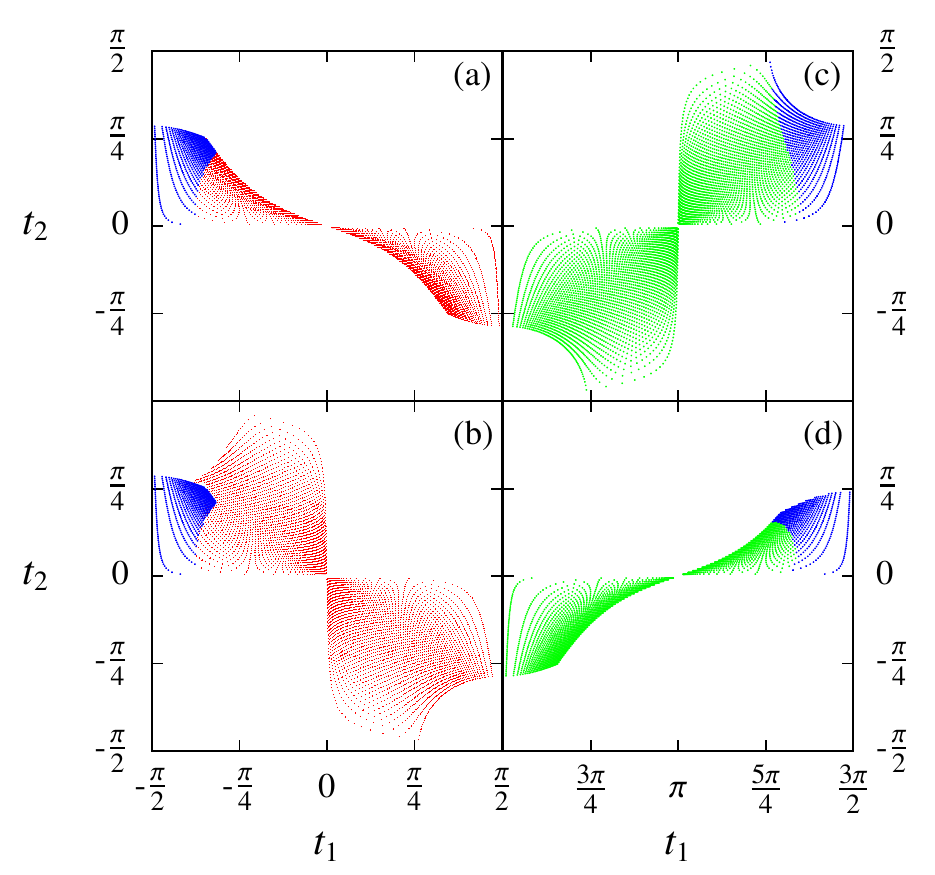}
\caption{Subsets of the EDS, labelled by symbol codes ${(f_1+;i_1-)}$. (a) ${(++;+-)}$. 
(b) ${(++;--)}$. (c) ${(-+;+-)}$. (d) ${(-+;--)}$.}
\label{EDS_E=0.03_class11}
\end{center}
\end{figure}

\begin{figure}[H]
\begin{center} 
\includegraphics[scale=1.0]{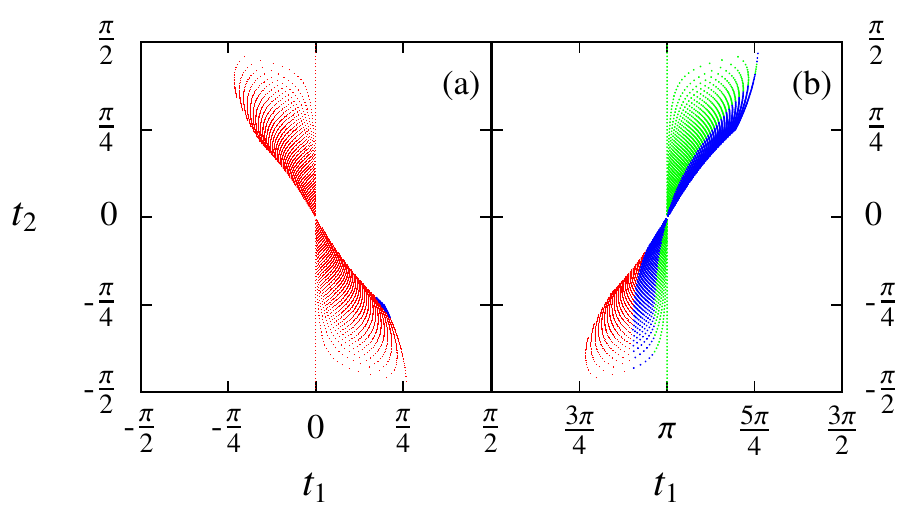}
\caption{Subsets of the EDS ${(f_1+;i_1+)}$. (a) ${(++;-+)}$. (b) ${(-+;++)}$.}
\label{EDS_E=0.03_class12}
\end{center}
\end{figure}

Figures \ref{EDS_E=0.03_class11} and \ref{EDS_E=0.03_class12} 
show the different parts of the EDS for the two subclasses
${(f_1+;i_1-)}$ and ${(f_1+;i_1+)}$.
Different types of trajectories on the EDS are represented by 
different colors: red for type 1$_2$, green for type 1$_1$ and blue for type 2.

The interpretation of these results is quite clear. 
The fate of trajectories in each of these subclasses 
is apparently controlled to a large extent by the dynamics 
associated with the first saddle plane. 
Classifying the results according to the symbols in this plane,
trajectories in the subsubclass ${(++;+-)} \cup {(++;--)}$ exhibit mainly 
type 1$_2$ behavior, as expected for trajectories which 
either cross the first saddle direction from $Q_1 < 0$ to $Q_1 >0$ 
or from trajectories which remain on the half of the plane $(Q_1,P_1)$ 
with $Q_1>0$. 
Referring to Figure \ref{poteigenvectors}, we see that 
these trajectories are those for which the bond $r_2$ is anticipated to  break. 
In the same manner, trajectories in the subsubclass ${(-+;--)} \cup {(-+;+-)}$ are
mainly type 1$_1$.

An important observation is the fact that type 2 trajectories (those exhibiting double bond fission) 
are not distributed randomly on these parts of the EDS; 
rather, type 2 trajectories actually occur at the \emph{boundary} between type 1$_1$ and 1$_2$
(cf.\ refs  \onlinecite{Andrews84,Grice87}).

\begin{figure}[H]
\begin{center} 
\includegraphics[scale=1.0]{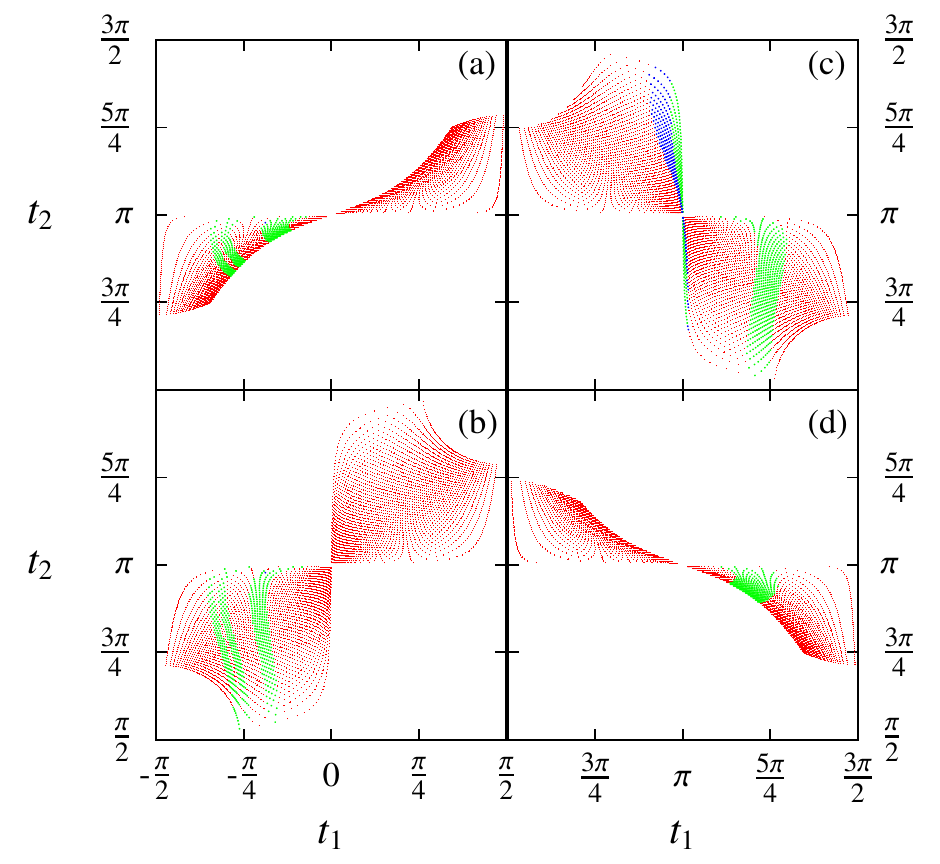}
\caption{Subsets of EDS ${(f_1-;i_1+)}$. (a) ${(+-;++)}$. (b) ${(+-;-+)}$. 
(c) ${(--;++)}$. (d) ${(--;-+)}$.}
\label{EDS_E=0.03_class21}
\end{center}
\end{figure}

\begin{figure}[H]
\begin{center} 
\includegraphics[scale=1.0]{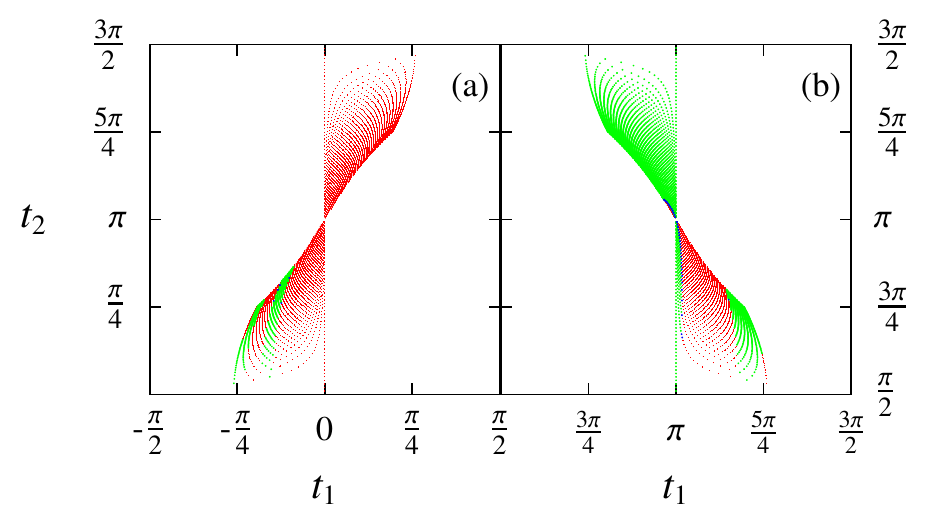}
\caption{Subsets of EDS ${(f_1-;i_1-)}$. (a) ${(+-;--)}$. (b) ${(--;+-)}$.}
\label{EDS_E=0.03_class22}
\end{center}
\end{figure}

We now turn to the second class of trajectory, consisting of sets 
${(f_1-;i_1+)} \cup {(f_1-;i_1-)}$. 
Again, this class splits into two subclasses ${(f_1-;i_1+)}$ and ${(f_1-;i_1-)}$,
and those subclasses split again according to the dynamics 
in the first saddle plane. 
Figures \ref{EDS_E=0.03_class21} and \ref{EDS_E=0.03_class22} 
show the results for these two subclasses. 
The situation is now very different from the previous case. 
All the trajectories belonging to these two subclasses 
enter the well region. 
The subsequent dynamics is then much more complex, 
and we are unable to predict which bond will be broken for a trajectory initiated 
on this portion of the EDS.

\begin{figure}[H]
\begin{center} 
\includegraphics[scale=1.0]{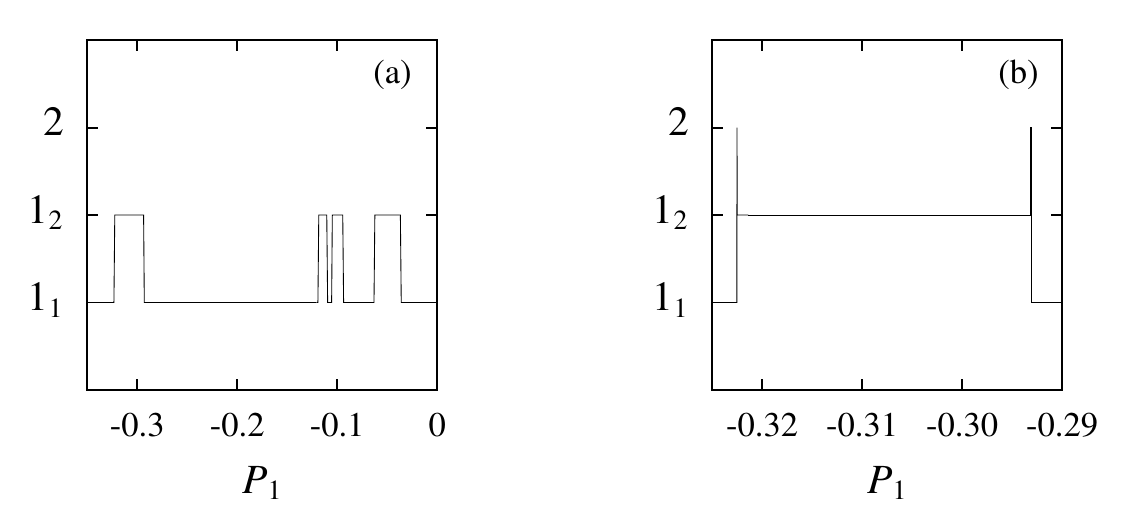}
\caption{(a) Trajectory type as a function of 
location along the 1D cut of the EDS: $Q_1=0.2$, $P_1 =-0.35$--$0.0$,
$E=0.03$. 
(b) Magnified segment showing the appearance of type 2 trajectories 
at the boundary between type 1$_1$ and 1$_2$.}
\label{EDS_cut_E=0.03}
\end{center}
\end{figure}

It should nevertheless be emphasized that the distribution of the 
different types of trajectory within these parts of the EDS is by
no means random and that we readily recognize 
a pattern of red and green `stripes', corresponding 
to alternating single bond breakages.
The interesting question then arises as to what happens at the 
boundary between types 1$_1$ and 1$_2$. In order to answer this 
question we sampled a line of initial conditions on the EDS 
which intersects these alternation of 
red and green strips. The results are
shown in figure \ref{EDS_cut_E=0.03}. 
Figure \ref{EDS_cut_E=0.03}(a) shows trajectory type versus 
the $P_1$ coordinate (used to sample the line of initial conditions);
the results indicate an alternation of the two types 
1$_1$ and 1$_2$. On the scale of Figure \ref{EDS_cut_E=0.03}(a),
the transition between one type to the other seems to be abrupt;
however, sampling more densely we get the results shown on Figure 
\ref{EDS_cut_E=0.03}(b), where now type 2 behaviour appears at the boundary between 
type 1$_1$ and 1$_2$. 

The interpretation of this result is again very simple. 
As the energy at which we sample the EDS is greater than the energy of the 
EP$_4$ saddle, trajectories in principle have enough energy to 
pass over the index 2 saddle and break 2 bonds. By continuity,
as we pass from the set of initial conditions in which the single bond $r_1$ breaks
to the set where  bond $r_2$ breaks, we traverse 
the region of the EDS where both bonds break (cf.\ again the work
of Chesnavich, refs \onlinecite{Andrews84,Grice87}).

\subsubsection{Energy below the index 2 saddle and `bond healing'}

We now turn to the behavior of trajectories initiated 
on the EDS at energies below the energy of the equilibrium EP$_4$. 

As discussed in the previous subsection, 
the results are easier to understand if we consider separately 
the part of the EDS consisting of trajectories which immediately escape from
the well region and the part for which trajectories enter the well.

\begin{figure}[H]
\begin{center} 
\includegraphics[scale=1.0]{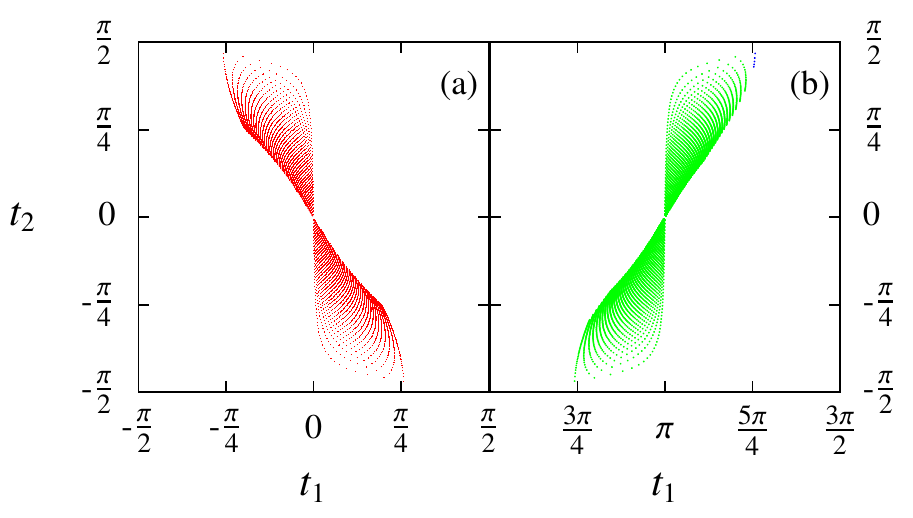}
\caption{EDS subsets ${(f_1+;i_1-)}$. (a) ${(++;+-)}$. (b) ${(-+;--)}$.}
\label{EDS_E=-0.03_class11}
\end{center}
\end{figure}

\begin{figure}[H]
\begin{center} 
\includegraphics[scale=1.0]{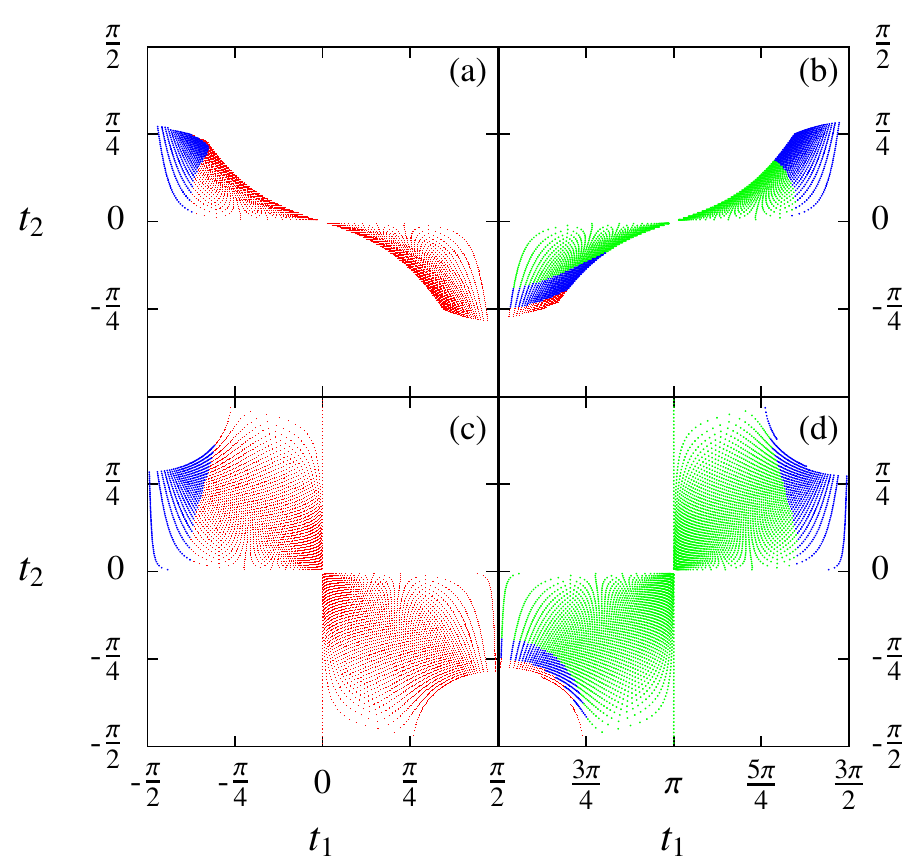}
\caption{EDS subsets ${(f_1+;i_1+)}$. (a) ${(++;-+)}$. (b) ${(-+;++)}$. 
(c) ${(++;++)}$. (d) ${(-+;-+)}$.}
\label{EDS_E=-0.03_class12}
\end{center}
\end{figure}

As before, the first class of trajectories is ${(f_1+;i_1-)} \cup {(f_1+;i_1+)}$.
As for the higher energy case, type 2 trajectories appear at the 
boundary between types 1$_1$ and 1$_2$ and we obtain mainly type 1$_1$ 
behavior for subsets having symbols $(-;+)$ and $(-;-)$ for motion in saddle
plane 1 and conversely mainly type 1$_2$ for subsets having symbols $(+;+)$ and $(+;-)$.
These two subclasses are represented on figures \ref{EDS_E=-0.03_class11} 
and \ref{EDS_E=-0.03_class12}

\begin{figure}[H]
\begin{center} 
\includegraphics[scale=1.0]{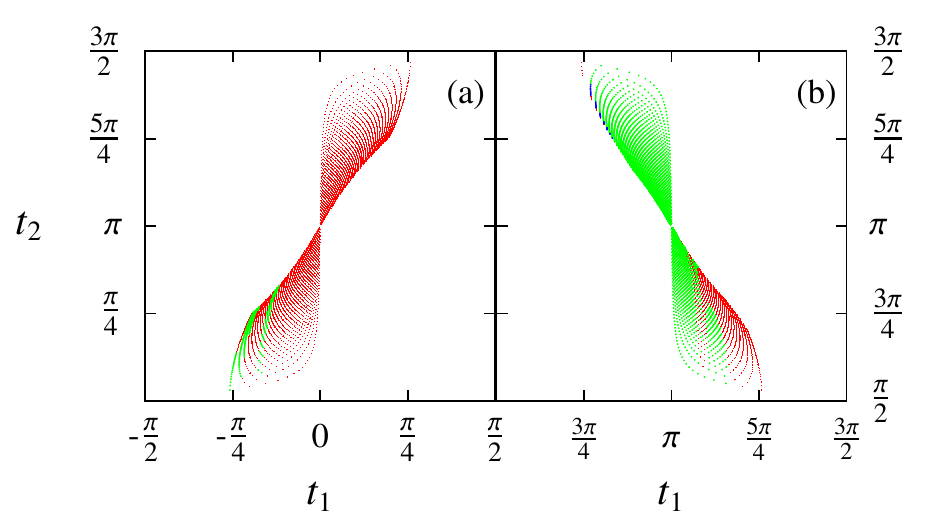}
\caption{EDS subsets ${(f_1-;i_1+)}$. (a) ${(+-;++)}$. (b) ${(--;-+)}$.}
\label{EDS_E=-0.03_class21}
\end{center}
\end{figure}

\begin{figure}[H]
\begin{center} 
\includegraphics[scale=1.0]{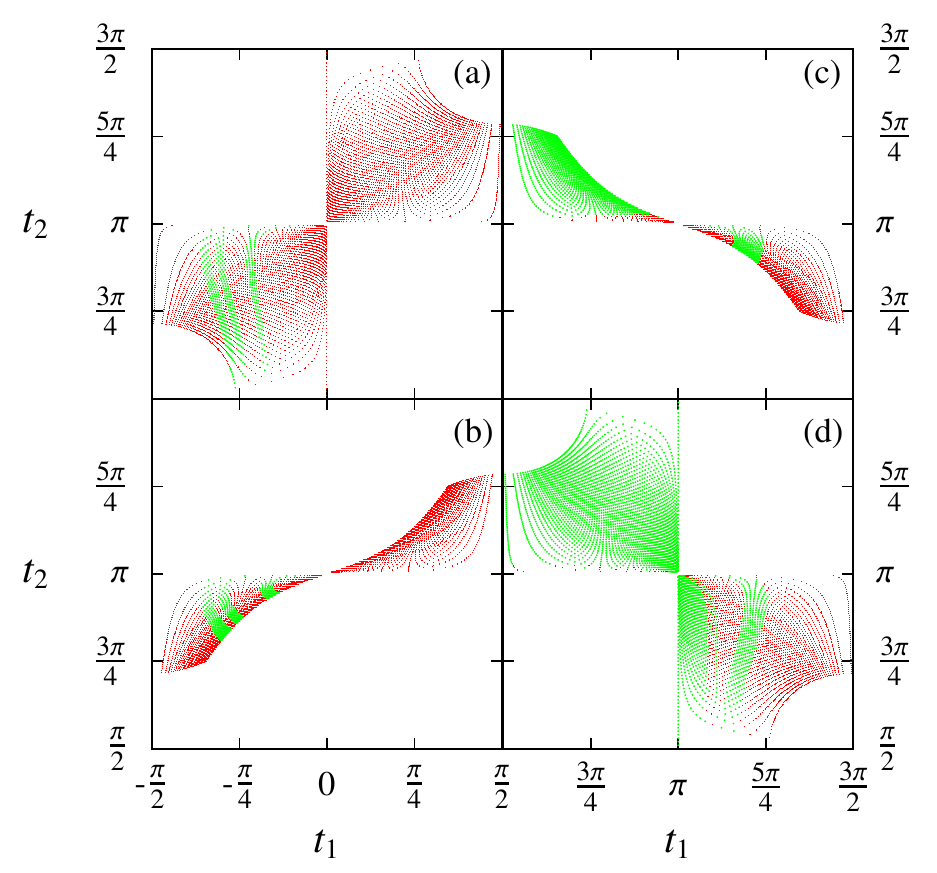}
\caption{EDS subsets ${(f_1-;i_1-)}$. (a) ${(+-;+-)}$. (b) ${(+-;--)}$. 
(c) ${(--;+-)}$. (d) ${(--;--)}$.}
\label{EDS_E=-0.03_class22}
\end{center}
\end{figure}

\begin{figure}[H]
\begin{center} 
\includegraphics[scale=0.5]{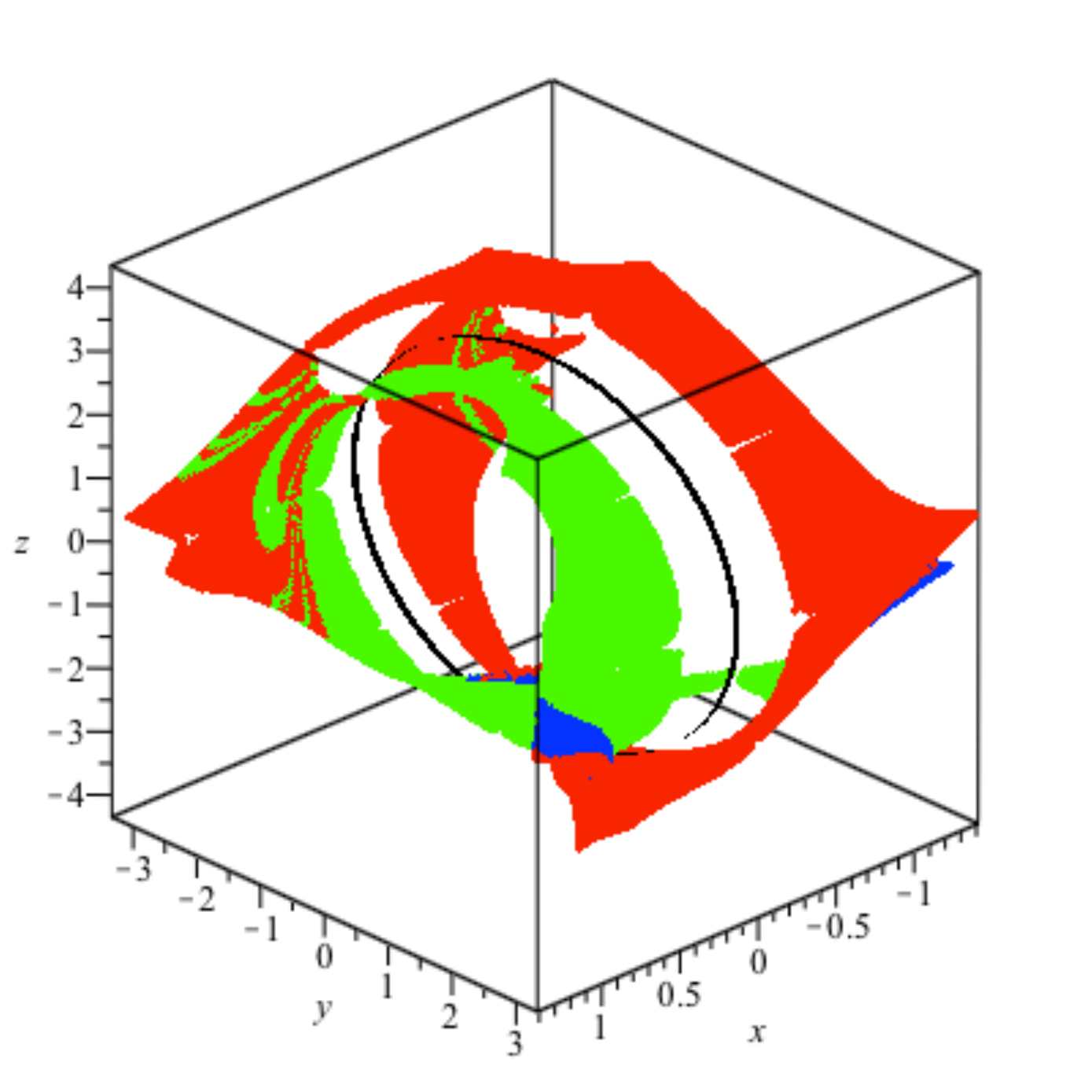}
\caption{Global representation of the EDS using the toroidal representation.
Energy $E=-0.03$.}
\label{toroidal_EDS_E=-0.03}
\end{center}
\end{figure}

The second class, as for the case of energies above that of the saddle, 
show a complex dynamics with alternation of type 1$_1$ and 1$_2$ trajectories. The
two subclasses of this class are represented on figure \ref{EDS_E=-0.03_class21} 
and \ref{EDS_E=-0.03_class22}.
Again, our interest will focus on what happens at the boundary. 
If we sample along a line of initial conditions we see an abrupt change between these types;
however, even if we sample on a very fine grid we do \emph{not} see 
type 2 trajectories at the boundary. Figure \ref{toroidal_EDS_E=-0.03} shows the global representation of
the EDS in the toroidal representation at energy $E=-0.03$.

\begin{figure}[H]
\begin{center} 
\includegraphics[scale=1.0]{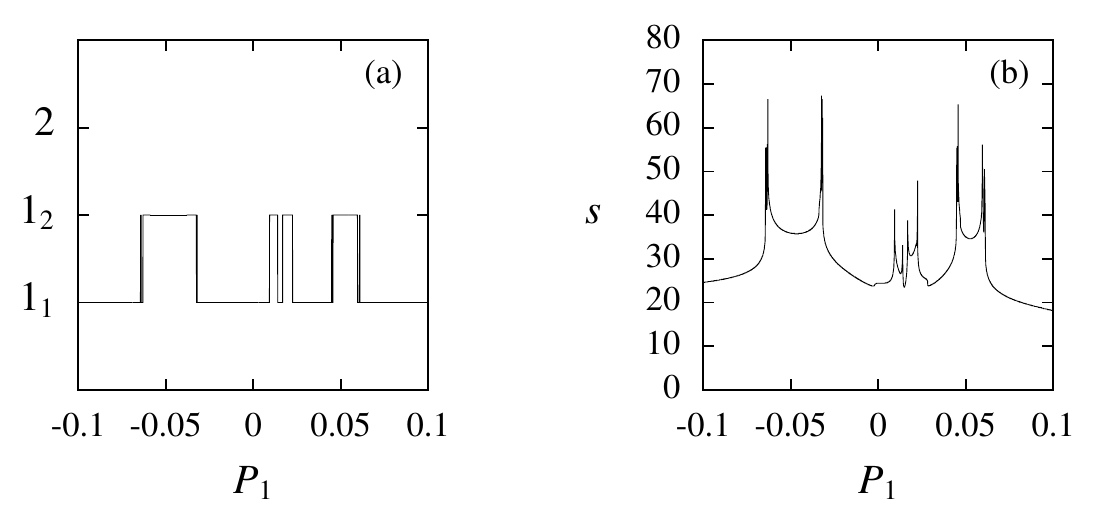}
\caption{(a)  Trajectory type as a function of 
location along the 1D cut of the EDS: $Q_1=0, P_1 =-0.1$--$0.1$,
$E = -0.03$. 
(b) Trajectory exit time  $s$ in time units along the cut.}
\label{EDS_cut_E=-0.03}
\end{center}
\end{figure}

In Figure \ref{EDS_cut_E=-0.03} we examine the trajectory exit time  
(the time it takes for a trajectory to actually dissociate, either along bond $r_1$ or $r_2$) 
along the line of initial conditions.
Figures \ref{EDS_cut_E=-0.03}(a), (b) show the trajectory type and corresponding exit
time, respectively, along the sampling line. 
At the boundary between types 1$_1$ and 1$_2$,
the exit time becomes very large, and appears to diverge as 
the sampling density is increased (results not shown here). 
We conclude that, at the boundary between type 1$_1$ and 1$_2$ trajectories, 
there exists a set of measure zero for which the exit time is infinite. 
In other words, these are trajectories that are trapped 
in the well region for $t \to +\infty$. The interpretation
of this result is quite familiar \cite{Pechukas81}: 
trajectories do not have enough energy to overcome 
the barrier for double bond breakage so that, in order to dissociate, 
either the $r_1$ bond or the $r_2$ bond must break.
Between these 2 possibilities, we have 
trajectories that take an infinite time to `decide'
between dissociation channels, and so remain trapped in the well.
(Note that, although we initiate trajectories in the vicinity of
index 2 saddle, trajectories can dissociate either by passing 
close to the index 2 saddle or by passing through the
`usual' transition state associated with one of the index 1 saddles. 
In the present work, we do not explore the interesting and 
important dynamical interplay between index 1 and index 2 saddles (cf. 
ref.\ \onlinecite{Harding12}), nor do we investigate the possibility of defining
a `global' dividing surface encompassing both types of saddle \cite{Harding12}.)

Examination of those trajectories which exhibit a large exit time 
suggests a connection between the form of these trajectories in 
configuration space and the so-called \emph{bond healing} phenomenon \cite{Oliveira94,Ghosh10,Paturej11}. 

\begin{figure}[H]
\begin{center} 
\includegraphics[scale=1.0]{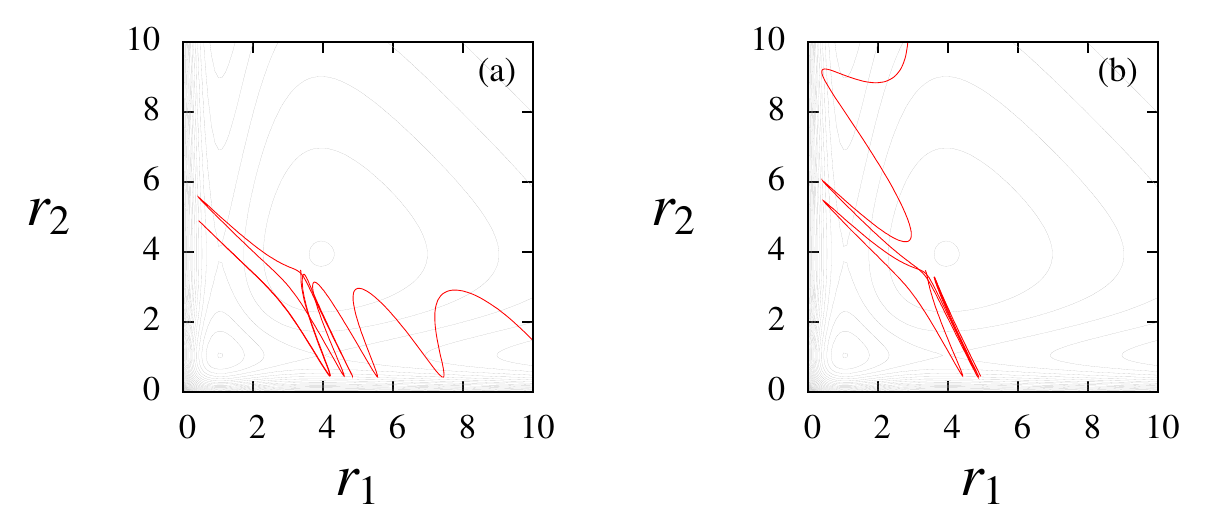}
\caption{Examples of trajectories exhibiting the phenomenon of `bond healing'.}
\label{plotselfhealing}
\end{center}
\end{figure}

Figure \ref{plotselfhealing} 
shows two examples of trajectories having a large exit time.  
The phenomenon of bond healing (as we interpret previous discussions 
\cite{Oliveira94,Ghosh10,Paturej11} of the concept) 
refers to the incipient breakage of one bond of the chain, followed 
by a recombination or `healing' of the bond rather than dissociation. 
The trajectories shown in 
Figure \ref{plotselfhealing} exhibit precisely this behavior; 
the trajectories pass well outside the nominal transition state (index 1 saddle)
before returning to the well region before ultimately dissociating.
In fact, these trajectories exhibit what might be called alternating bond healing, 
where one bond almost breaks and then reforms and then the other bond
breaks and reforms, and so on and recombined. 
Such trajectories oscillate in an anti-diagonal direction in the $(r_1,r_2)$ plane.

An important task for future investigation 
is the identification and computation of invariant 
phase space objects (periodic orbits, NHIMs) 
which are responsible for trapping trajectories in their vicinity, 
leading to divergent exit times.

\newpage

\section{Conclusions and perspectives}
\label{conclusec}

In this paper we have investigated the 
fragmentation dynamics of an atomic chain under tensile stress.
We have analyzed the location and types (indices) of equilibria for
the general $n$-particle chain, and have noted the importance of 
saddle points with index $> 1$. 

For an $n=2$-particle chain under tensile stress the index 2 saddle is of
key significance for the dynamics.
Building upon previous work, we apply normal form theory to 
analyze phase space structure and dynamics in a neighborhood of the index 2 saddle.
We define a phase dividing surface (DS) that enables us to 
classify trajectories passing through a neighborhood of the 
saddle point using the values of the integrals associated with the normal form.  We 
also generalize our definition of the dividing surface and define an
\emph{extended dividing surface} (EDS), which is used to sample and classify trajectories
that pass through a phase space neighborhood of the index 2 saddle at total energies
less than that of the saddle.

Classical trajectory simulations are used to study
single versus double bond breakage for the $n=2$ chain under tension.  
Initial conditions for trajectories are obtained by sampling the EDS at constant energy.
We sample trajectories at fixed energies both above and below the energy of the saddle.  
The fate of trajectories (single versus double bond breakage) is 
explored as a function of the location of the initial condition on the EDS, and
connection made to the work of Chesnavich on collision-induced dissociation \cite{Andrews84,Grice87}.
A significant finding is that we can readily identify
trajectories that exhibit bond \emph{healing} \cite{Oliveira94,Ghosh10,Paturej11}; 
such trajectories pass outside the nominal transition state for 
single bond dissociation, but return to the well region possibly several times before
ultimately dissociating. 
Identification of the invariant phase space structures associated with 
trapped trajectories is a topic for future investigation.

\acknowledgments

FM, PC, and SW  acknowledge the support of the  Office of Naval Research (Grant No.~N00014-01-1-0769) and the Leverhulme Trust.

\newpage

\appendix

\section{Sampling the extended dividing surface for index $k$ saddles}
\label{appendixsampling}

The concept of the extended dividing surface (EDS) (as introduced in
Section \ref{normformsec}) is of essential importance for our study. 
The precise definition of this phase space object 
is given in terms of the normal form Hamiltonian. 
A procedure for sampling phase space points on the dividing surface
associated with index-2 saddles was introduced and implemented 
in Ref.\ \onlinecite{Collins11}. 
In this Appendix we present another method of sampling 
the Normal Form EDS which extends quite naturally to the case 
of index $k$ saddles where $k \geq 2$.

We start by recalling the definition of the saddle actions in terms of the $(Q_i,P_i)$ coordinates:
\begin{equation}
I_i = \frac{1}{2} (P_i^2 - Q_i^2), \:\: i=1,\dots,k.
\label{saddleactions}
\end{equation}
As the actions $I_i$ are constants of the motion, we have in each saddle plane the equation:
\begin{equation}
\frac{P_i^2}{2I_i} - \frac{Q_i^2}{2I_i} = 1,
\label{hyperbolaequat}
\end{equation}
which is just the equation of a doubly-branched hyperbola. 
There is then a very simple parametrization of the hyperbola in terms of one parameter 
$t_i$, where the form of the parametrization
depends on the sign of $I_i$.
For the case $I_i > 0$ we have:
\begin{subequations} 
\label{hyperbolaparam_Ipos}
\begin{align}
Q_i & = \sqrt{2I_i} \tan(t_i)
\label{hyperbolaparam_Ipos_Q} \\
P_i & = \frac{\sqrt{2I_i}} {\cos(t_i)}, \:\: 
t_i \in \left ]\frac{-\pi}{2};  \frac{\pi}{2}\right [ \, \cup \,
\left ]\frac{\pi}{2};\frac{3\pi}{2}\right[.
\label{hyperbolaparam_Ipos_P}
\end{align}
\end{subequations}
while for $I_i<0$ we take:
\begin{subequations} 
\label{hyperbolaparam_Ineg}
\begin{align}
Q_i & = \frac{\sqrt{2|I_i|}} {\cos(t_i)} 
\label{hyperbolaparam_Ineg_Q}\\
P_i & = \sqrt{2|I_i|} \tan(t_i), \:\: t_i \in \left]\frac{-\pi}{2}; \frac{\pi}{2}\right[ \, \cup \,
\left] \frac{\pi}{2}; \frac{3\pi}{2} \right[.
\label{hyperbolaparam_Ineg_P}
\end{align}
\end{subequations}

In ref \onlinecite{Collins11} 
the dividing surface was defined as a codimension one surface within 
the energy surface consisting of phase points 
which minimized the distance from the origin, where the 
distance $\overline{D}$ was defined in configuration ($Q_i$) space:
\begin{equation}
\overline{D} = \frac{1} {2} \sum_{i=I}^k Q_i^2.
\label{distanceold}
\end{equation}
The use of the distance $\overline{D}$ in ref.\ \onlinecite{Collins11}   
was appropriate because the two parameters $R$ and $\theta$ 
used to parametrize the DS specified the location of phase points in the 
$(Q_1,Q_2)$ plane (index 2 case).   Conservation of energy together 
with the minimum distance condition then served to define the corresponding
momenta $(P_1,P_2)$.

If we use the parametrization of the hyperbolas of eq.\
\eqref{hyperbolaparam_Ipos} and \eqref{hyperbolaparam_Ineg} 
it is more natural to introduce a new distance defined in the  full phase space
(cf. eq.\ \eqref{defdistance}):
\begin{equation}
D = \frac{1} {4} \sum_{i=I}^k \left( Q_i^2 + P_i^2 \right).
\label{distancenew}
\end{equation}
The derivatives with respect to 
time of the two distances $\overline{D}$ and $D$ 
are in fact \emph{identical}, so that 
for both cases the minimum distance condition is:
\begin{equation}
\dot{D} = \sum_{i=I}^k \Lambda_i Q_i P_i=0.
\label{distancenew_2}
\end{equation}
Independent of the sign of $I_i$, 
the following equation then defines the constant energy EDS 
($K(I_1,\dots,I_n) = E$):
\begin{equation}
\sum_{i=I}^k 2 \Lambda_i |I_i| \; \frac{\tan(t_i)}{\cos(t_i)} =0.
\label{mindistequa}
\end{equation}
In terms of the variables $X_i \equiv {\tan(t_i)}/{\cos(t_i)}$, we have:
\begin{equation}
\sum_{i=I}^k 2 \Lambda_i |I_i| X_i=0.
\label{mindistequaX}
\end{equation}

As noted above, the EDS is composed of $4^k$ parts 
for the case of an index $k$ saddle. 
For each saddle plane, the hyperbola has four branches, 
two with $I_i > 0$ and two with $I_i < 0$.
The four branches are in 1:1 correspondence with the 
symbolic notation introduced in Section \ref{normformsec}:
\begin{subequations}
\begin{align}
 I_i>0 \; \text{and} \; t_i \in \left]\frac{-\pi}{2};\frac{\pi}{2}\right[ \, & \longleftrightarrow \, (+;-) \\
 I_i>0 \; \text{and} \; t_i \in \left]\frac{\pi}{2};\frac{3\pi}{2}\right[ \, & \longleftrightarrow \, (-;+) \\
 I_i<0 \; \text{and} \; t_i \in \left]\frac{-\pi}{2};\frac{\pi}{2}\right[ \, & \longleftrightarrow \, (+;+) \\
 I_i<0 \; \text{and} \; t_i \in \left]\frac{\pi}{2};\frac{3\pi}{2}\right[ \, & \longleftrightarrow \, (-;-).
\end{align}
\end{subequations}

From the equation $X_i = {\tan(t_i)}/{\cos(t_i)}$ we obtain:
\begin{equation}
X_i^2 = \left( \frac{\tan(t_i)}{\cos(t_i)} \right)^2 = \frac{1}{\cos^4(t_i)} - \frac{1}{\cos^2(t_i)}.
\label{Xt-relation}
\end{equation}
which yields
\begin{equation}
z_i^2 - z_i -X_i^2 = 0
\label{zsecondorderequa}
\end{equation}
in terms of the new variable $z_i={1}/{\cos^2(t_i)}$.
The discriminant $\Delta=1+4X_i^2$ of eq.\ \eqref{zsecondorderequa} 
is always positive and there are two roots:
\begin{equation}
(z_i)_{\pm} = \frac{1\pm \sqrt{\Delta}}{2}.
\label{zroots}
\end{equation}
However, $z$ is also a square ($z= {\cos(t_i)}^{-2} $) 
and the $(z_i)_-$ solution is always negative or zero. 
So, the only possible solution is actually $(z_i)_+$, so we finally have for $t_i$ 
the expression:
\begin{equation}
t_i = \cos^{-1} \left[ \pm \sqrt{ \frac{2} {1 +\sqrt{\Delta} } } \, \right].
\label{troots}
\end{equation}
If we set $\alpha = \sqrt{ \frac{2} {1 +\sqrt{\Delta} } }$, 
we see that $0 \leq \alpha \leq 1$ and $-1 \leq -\alpha \leq 0$ and 
consequently $\cos^{-1}(\alpha) \in [0;\frac{\pi}{2}]$ and 
$\cos^{-1}(-\alpha) \in [\frac{\pi}{2}; \pi]$. 
It is straightforward to choose the appropriate root depending 
on which branch of the hyperbola we are working on 
and making the appropriate translation of the set $[0;\pi]$ to 
either $[-\frac{\pi}{2};\frac{\pi}{2}]$ or $[\frac{\pi}{2};\frac{3\pi}{2}]$.

To sample the EDS, we first  
have to fix the value of the energy $E$, 
which can be either positive or negative. 
We must therefore find a set of actions $I_i$ for which 
eq.\ \eqref{HequalE} holds. 
This can be done by choosing $n-1$ actions 
and solving numerically for the value of the last action $I_s$. 
This set of actions can be chosen in a systematic way. We 
can specify maximal or minimal values for each action depending
on the part of the EDS we are sampling and sample systematically 
from zero to the maximal or the minimal value of each actions. The
determination of these maximal or minimal values of the actions is 
related to the accuracy of the normal form Hamiltonian and 
its associated coordinates transforms  \eqref{changevar2}. The check
of the accuracy of the normal form Hamiltonian is an important topic and we discuss some aspects of this topic relevant to this work  in the next paragraph.

Normal form theory is procedure which enable us to construct a "simple"
Hamiltonian which approximates the dynamics of a given Hamiltonian in a
neighbourhood of an equilibrium point of this Hamiltonian. The outputs
of this procedure are a normal form Hamiltonian and two sets of coordinates
transformations which enable us to navigate between the two sets of
physical and normal form coordinates. The construction of the normal form Hamiltonian relies on a Taylor
expansion about an equilibrium point and a normalisation process 
which results in  the normal form being valid only in a certain neighbourhood 
of the equilibrium point under consideration. Therefore we expect the normal
form to ''break down'', i.e. not be an accurate approximation of the physical Hamiltonian, 
when we leave this neighbourhood. The
major question here is how to define this neighbourhood or, in other words, 
how to measure the accuracy of the normal form? For Hamiltonian
systems there is a quite natural measure of the accuracy of the normal form
which is the conservation of  energy. The idea used here is to compare the energy
provided by the normal form Hamiltonian with the energy provided by the initial
Hamiltonian and decide a
threshold for energy "disagreement" between the two:

\begin{equation}
|E_{ini}-E_{NF}| \le \epsilon.
\label{energydiff}
\end{equation}

\noindent
For phase space points in normal form coordinates we compute the energy with
the normal form Hamiltonian and use the backward transformations \eqref{changevar2}
provided by the normalisation process to obtain phase space points in physical coordinates
with which we can compute the  energy using the physical Hamiltonian. Starting
from phase space points "close" to the equilibrium point, and increasing the distance from this equilibrium
point gradually, we can construct a connected set of phase space points for which \eqref{energydiff}  holds, 
and which for the purpose of this work defines the neighbourhood of validity of the normal form.
This neighbourhood, denoted $\mathcal{L}_{\epsilon}$, depends on $\epsilon$. 
The parameter $\epsilon$ actually quantifies the "error threshold" in the normal form. For this work the value
of $\epsilon$ was taken as a factor
$10^{-3}$ times the physical energy of the energy surface on which the EDS was sampled.

We now return to the problem of computing the maximal or minimal values of the actions. We can
differentiate between the case of a saddle mode action and a bath mode action. For a bath mode with action $I_b$ 
the dynamics is confined to a circle of radius $r=\sqrt{2I_b}$ in this bath mode plane. 
To determine the maximal 
value of $I_b$ we set all other coordinates related to the other modes (bath and saddle modes) 
to zero so that the phase space
point is located at the origin of all the mode planes except the one we try to determine the maximum value of the action. For
a certain value of the action $I_b$ we calculate the energy $\epsilon$-agreement for samples all around the circle
of radius $r=\sqrt{2I_b}$. As we increase the action $I_b$ the radius of the circle will increase and we will find some points
on the circle which do not belong to $\mathcal{L}_{\epsilon}$. So the maximal value of the action $I_b$ will be the one
corresponding to the last circle for which all the points on the circle belong to $\mathcal{L}_{\epsilon}$. For the action
corresponding to a saddle mode we can have two cases depending on which part of the EDS we are sampling but the
procedure is quite identical for the two cases. For a saddle mode of fixed action $I_s$ on a certain part of the EDS
the dynamics is constrained on one branch of a hyperbola. As for the case of a bath mode we set all coordinates related
to the other modes to zero. On the branch of the hyperbola the minimal distance from the origin in the saddle mode plane is at $t=0$ or $\pi$
depending on the part of the EDS. So if we fix $t$ to 0 or $\pi$ and increase ($I_s>0$) or decrease ($I_s<0$) we will
pass  a point for which the energy $\epsilon$-agreement between normal form Hamiltonian energy and initial energy does not
holds which means that we have left $\mathcal{L}_{\epsilon}$. This determines the maximal or minimal value of the action $I_s$.

Having selected a particular set of bath mode and saddle actions 
we sample points which belong to the EDS by assigning values to
the the variables that are complementary to the actions.
For each of the bath mode planes we sample the angle variable
conjugate to the actions. If a bath mode action is $I_b$,
a conserved quantity, we have $I_b = \frac{1}{2}(q_b^2+p_b^2)$.
The angle $\theta$, $0 \leq \theta \leq 2 \pi$,
parameterizes a circle of radius  $r=\sqrt{2I_b}$ in the $(q_b,p_b)$ plane. 

For the saddle planes, we need 
to solve the minimum distance equation to sample points on the EDS. 
Equation \eqref{mindistequaX} provides us with a very simple 
relation among the variables $X_i$, $i=1,\dots,k$. 
By fixing all of those variables except  $X_1$ and $X_2$, for example,
equation \eqref{mindistequaX} describes the equation of a straight line 
with slope $-\frac{\Lambda_2 |I_2|}{\Lambda_1 |I_1|}$ and 
$Y$-intercept 
$\sum_{j=3}^k  -\frac{\Lambda_j |I_j|}{\Lambda_1 |I_1|} X_j$. 
For the case of only two saddle planes, $k=2$, it reduces to 
the equation of a straight line passing through the origin. 
After determining the $X_i$, we compute the $t_i$ as 
explained above and finally the $(Q_i,P_i)$.

\def\cprime{$'$}

\newpage



\newpage

\newpage

\newpage

\newpage

\newpage

\newpage

\newpage

\newpage

\newpage

\newpage

\newpage

\newpage

\newpage

\newpage

\newpage

\newpage

\newpage

\newpage

\newpage

\end{document}